\documentclass[preprintnumbers,article,amsmath,amssymb,floatfix,10pt,prd,onecolumn,superscriptaddress,nofootinbib]{revtex4-2}

\usepackage{cancel, ulem}\usepackage{adjustbox}
\usepackage{titlesec}
\usepackage[toc]{appendix}
\usepackage[font=small,labelfont=bf,justification=justified,singlelinecheck=false]{caption}
\titleformat*{\section}{\LARGE\bfseries}
\titleformat*{\subsection}{\Large\bfseries}
\titleformat*{\subsubsection}{\large\bfseries}
\titleformat*{\paragraph}{\large\bfseries}
\titleformat*{\subparagraph}{\large\bfseries}
\usepackage{bm}
\usepackage{amsfonts}
\usepackage{latexsym}
\usepackage[utf8]{inputenc}
\usepackage{graphicx}
\usepackage{amsmath}
\usepackage{palatino}
\usepackage{mathpazo}
\usepackage{textcomp}
\usepackage{float}
\usepackage{booktabs}
\usepackage{dcolumn}
\usepackage{ragged2e}
\usepackage{hyperref}
\hypersetup{colorlinks,citecolor=blue}
\hypersetup{colorlinks=true,linkcolor=blue,filecolor=magenta,urlcolor=blue}
\usepackage{amsthm}
\usepackage{xcolor}
\usepackage{orcidlink}
\usepackage{epsfig}
\usepackage{commath}
\usepackage{cancel, ulem}
\usepackage{subcaption}
\usepackage{caption}
\usepackage{multirow}
\newcommand{\m}{\mathring}
\newcommand{\be}{\begin{equation}}
\newcommand{\ee}{\end{equation}}
\newcommand{\bea}{\begin{eqnarray}}
\newcommand{\eea}{\end{eqnarray}}
\newcommand{\eeas}{\end{eqnarray*}}
\newcommand{\beas}{\begin{eqnarray*}}

\def\jnl@style{\it}
\def\aaref@jnl#1{{\jnl@style#1}}

\def\aaref@jnl#1{{\jnl@style#1}}

\def\aj{\aaref@jnl{AJ}}                   
\def\apj{\aaref@jnl{ApJ}}                 
\def\apjl{\aaref@jnl{ApJ}}                
\def\apjs{\aaref@jnl{ApJS}}               
\def\apss{\aaref@jnl{Ap\&SS}}             
\def\aap{\aaref@jnl{A\&A}}                
\def\aapr{\aaref@jnl{A\&A~Rev.}}          
\def\aaps{\aaref@jnl{A\&AS}}              
\def\mnras{\aaref@jnl{Mon.~Not.~Roy.~Astron.~Soc.}}             
\def\prd{\aaref@jnl{Phys.~Rev.~D}}        
\def\prc{\aaref@jnl{Phys.~Rev.~C}}  
\def\prl{\aaref@jnl{Phys.~Rev.~Lett.}}    
\def\qjras{\aaref@jnl{QJRAS}}             
\def\skytel{\aaref@jnl{S\&T}}             
\def\ssr{\aaref@jnl{Space~Sci.~Rev.}}     
\def\zap{\aaref@jnl{ZAp}}                 
\def\nat{\aaref@jnl{Nature}}              
\def\aplett{\aaref@jnl{Astrophys.~Lett.}} 
\def\apspr{\aaref@jnl{Astrophys.~Space~Phys.~Res.}} 
\def\physrep{\aaref@jnl{Phys.~Rep.}}      
\def\physscr{\aaref@jnl{Phys.~Scr}}       
\def\commat{\aaref@jnl{Comm.~Math.~Phys.}}              
\def\science{\aaref@jnl{Science}}               
\def\cqg{\aaref@jnl{Classical Quant.~Grav.}}            
\def\jpcs{\aaref@jnl{JPCS}}                                     
\def\ijmpd{\aaref@jnl{Int.~J.~Mod.~Phys.~D}}                    
\def\grg{\aaref@jnl{Gen.~Relat.~Gravit.}}               
\def\rpp{\aaref@jnl{Rep.~Prog.~Phys.}}          
\def\npa{\aaref@jnl{Nucl.~Phys.~A}}        
\def\lrr{\aaref@jnl{Living Rev.~Rel.}}                   
\def\jcap{\aaref@jnl{J.~Cosmology Astropart.~Phys.}}    
\def\rmp{\aaref@jnl{Rev.~Mod.~Phys.}}   
\def\epjc{\aaref@jnl{Eur.~Phys.~J.~C}} 
\def\plb{\aaref@jnl{~Phy.~Lett.~B}} 
\def\mpla{\aaref@jnl{Mod.~Phy.~Lett.~A}} 
\def\arxiv{\aaref@jnl{arxiv.org}}

\allowdisplaybreaks[1]

\begin{document}
\title{Quintaverse-Phantiverse Cosmologies in $f(Q,C)$ Gravity: a Non-coincident Geometrical Origin of Cosmic Acceleration}
\author{Hamid Shabani\orcidlink{0000-0002-2309-3591}}
\email [Corresponding Author: ]{h.shabani@phys.usb.ac.ir}
\affiliation{Physics Department, Faculty of Sciences, University of Sistan and Baluchestan, Zahedan, Iran}
\author{Avik De\orcidlink{0000-0001-6475-3085}}
\email{avikde@um.edu.my}
\affiliation{Institute of Mathematical Sciences, Faculty of Science, Universiti Malaya, 50603 Kuala Lumpur, Malaysia}



\begin{abstract}
We investigate the asymptotic dark energy sectors of a quadratic form of $f(Q,C)$ Lagrangian. We reformulate the original equations of motion to obtain the equivalent dynamical system, for two non-trivial classes of affine connections allowed by symmetries of $f(Q,C)$ gravity. In these scenarios, the late-time accelerated expansion emerges as an intrinsic geometrical effect of the $f(Q,C)$ sector, without invoking an independent dark energy component. 
We find that an effective dark matter solution emerges for one class, whereas for the other class, the positivity of the matter density and the conservation of the momentum tensor cannot be simultaneously satisfied. This leads to the conclusion that there exists an effective dark matter solution for which the matter contribution is a combination of a pressureless matter as well as a stiff fluid.
Furthermore, in both cases there always exists a universal stable accelerated expansion solution. Besides, the matter solutions are saddle points for those values of the model parameters that ensure the conservation of the energy momentum. In addition, we introduce a parameterization in which the equation of state parameter at the fixed point corresponding to accelerated expansion is characterized by a free parameter that quantifies its deviation from $-1$. We thereby inspired to invent the terms ``Quintaverse'' and ``Phantiverse'', corresponding to quintessence-like and phantom-like attractor solutions. Also, we numerically illustrate the evolution of different cosmological quantities and the phase-space diagrams. Particularly, the behavior of the Universe using two kinematic toolkits, namely, the cosmographic and statefinder variables, is discussed.
\end{abstract}

\maketitle
\tableofcontents
\section{Introduction}\label{formulation_intro}

Observational cosmology over the past three decades, supported by Type Ia Supernovae (SNe Ia) compilations, Cosmic Microwave Background (CMB) anisotropy spectra, and Baryon Acoustic Oscillations (BAO), has robustly established that the Universe is experiencing late-time accelerated expansion \cite{Riess1998, Perlmutter1999, aghanim2020}. Within the standard Einstein-Hilbert paradigm of General Relativity (GR), this cosmic acceleration is conventionally parameterized by the $\Lambda$-Cold Dark Matter ($\Lambda$CDM) model, where a constant vacuum energy density yields a fixed equation-of-state (EoS) parameter $w = -1$. Despite its remarkable success in fitting background data, the cosmological constant $\Lambda$ suffers from deep-seated theoretical conundrums, most prominently the fine-tuning and cosmic coincidence problems, as well as growing statistical tensions in key parameters such as $H_0$ and $\sigma_8$ \cite{Bull2016, DiValentino2021}. 

More recently, high-precision cosmological datasets, most notably the Data Release 1 (DR1) and subsequent observations from the Dark Energy Spectroscopic Instrument (DESI) combined with Planck CMB and standardized SNe Ia compilations, have sparked renewed debate by indicating a statistical preference for dynamical dark energy over a rigid cosmological constant \cite{Roy2024, DESI2024_BAO}. In particular, model-agnostic expansion history reconstructions and parameterizations favor an evolving dark energy sector whose EoS parameter $w(z)$ hovers near, and potentially crosses, the "phantom divide" line $w = -1$ at low redshifts \cite{Roy2024, Roy2025_cross}.

This observational signature directly underscores the concept of **Quintaverse–Phantiverse Cosmologies**. Originally conceptualized within the framework of GR via scalar fields, dark energy regimes are broadly categorized into two distinct domains:
\begin{itemize}
    \item \textbf{Quintessence Regime ($w > -1$):} Modeled in GR via canonical scalar fields with positive kinetic energy, yielding an EoS that remains strictly above the phantom barrier ($w \ge -1$) \cite{Caldwell1998, Zlatev1999}. While well-behaved UV-complete theories can be constructed in this domain, canonical scalar fields cannot cross $w = -1$ without violating energy conditions.
    \item \textbf{Phantom Regime ($w < -1$):} Achieving $w < -1$ within GR necessitates non-canonical kinetic terms with negative energy density (ghost fields) \cite{Caldwell2002}. However, standard GR phantom fields suffer from severe quantum pathologies, including vacuum instability, ghost-driven quantum breakdown, and eventual "Big Rip" finite-time cosmic singularities.
\end{itemize}

To circumvent the classical and quantum instabilities inherent to GR phantom scalar fields, modified theories of gravity offer a compelling alternative: the effective dark energy stress-energy tensor emerges purely from spacetime geometry \cite{Nojiri2017, Capozziello2011}, and the effective EoS parameter $w_{\text{eff}}$ can naturally inhabit either the quintessence-like domain (\textbf{Quintaverse}) or the phantom-like domain (\textbf{Phantiverse}) as stable, smooth asymptotic attractors without introducing physical ghost degrees of freedom or pathological fine-tuning.

Among geometrically modified gravity paradigms, symmetric teleparallel gravity replaces the curvature-based constructions with a flat, torsion-free spacetime where gravitational interactions are mediated entirely by non-metricity $Q_{\lambda\mu\nu}$ \cite{adak, Jimenez2018, Heisenberg2024}. The difference between the Levi-Civita Ricci scalar $\m R$ and the non-metricity scalar $Q$ defines a boundary term $C = \m R - Q = -\m \nabla_\alpha (Q^\alpha - \tilde{Q}^\alpha)$. Incorporating this term leads to $f(Q,C)$ gravity, a recent unified framework that subsumes both $f(Q)$ and $f(\m R)$ gravities while maintaining complete gauge invariance under coordinate transformations \cite{als/fQC, Capozziello2023a, andro:fQB}. Some of the major background cosmological implications of the $f(Q,C)$ gravity can be found in \cite{Artola, Shabani:2026, shabani1:2025, Gadbail2023, Lohakare, Ghulam:2025, Usman:2024, Ayuso:, Samaddar:2025} and the references therein. The scalar perturbation and density contrast evolution in this gravity theory was recently published, paving the way for growth analysis and further observational contact with the theory \cite{Ganesh:2025}.

In a spatially flat Friedmann–Robertson–Walker (FLRW) spacetime, symmetric teleparallel affine connections allow non-trivial gauge choices specified by a dynamical connection function $\gamma(t)$ \cite{FLRW/connection, ad/bianchi}. The primary objective of this paper is to explore the cosmological viability of a generic quadratic model, $f(Q,C) = \eta_1 Q^2 + \eta_2 C^2 + \eta_3 CQ$, across two non-trivial connection classes (Class $II$ and Class $III$). Utilizing a novel $\epsilon$-parameterization ($w_{\text{eff}} = -1 + \epsilon$), we examine whether Quintaverse ($\epsilon > 0$) and Phantiverse ($\epsilon < 0$) behaviors can arise naturally as purely geometric late-time attractors.

The structure of this paper is organized as follows:\\
In Sect.~\ref{formulation}, we review the foundational formalism of $f(Q,C)$ gravity and introduce a particular quadratic model, whose physical implications for the background cosmological expansion are explored. 
In Sect.~\ref{type2}, we investigate the cosmological evolution of the Universe with the connection class $II$. We systematically examine different aspects of the underlying theory in several subsections. We also introduce the $\epsilon$-parameterization, which provides a parametrization of the effective {\sf EoS} parameter of the Universe during the {\sf AE} epoch. 
In Sect.~\ref{type3}, we repeat the analysis for Connection Class $III$, highlighting key dynamical differences such as matter-geometry couplings. 
In Sect.~\ref{numerical}, we present numerical reconstructions of background cosmological quantities, cosmographic parameters, and statefinder diagnostics. Finally, Sect.~\ref{conclusion} summarizes our principal findings.

\section{Basic formulation of $f(Q,C)$ gravity}\label{formulation}
In the framework of symmetric teleparallel theories of gravity, the affine connection is taken to be torsion-free and curvature-free, while the gravitational interaction is entirely encoded in the non-metricity of spacetime. The latter is quantified by the failure of the metric tensor to be covariantly conserved, which is expressed through the non-metricity tensor
\begin{equation} \label{Q tensor}
Q_{\lambda\mu\nu} = \nabla_\lambda g_{\mu\nu}=\partial_\lambda g_{\mu\nu}-\Gamma^{\beta}_{\,\,\,\mu\lambda}g_{\beta\nu}-\Gamma^{\beta}_{\,\,\,\nu\lambda}g_{\beta\mu}\neq 0,
\end{equation}
signalling that lengths and angles are not preserved under parallel transport. In this way, non-metricity plays the central role in mediating gravity, similar to curvature in General Relativity or torsion in metric teleparallel gravity.

A generic affine connection $\Gamma$ can be decomposed into the Levi-Civita connection $\mathring \Gamma$ and the disformation tensor $L^\lambda{}_{\mu\nu}$ as
\begin{equation} \label{connc}
\Gamma^\lambda{}_{\mu\nu} := \mathring{\Gamma}^\lambda{}_{\mu\nu}+L^\lambda{}_{\mu\nu},
\end{equation}
where the disformation encodes the pure effects of non-metricity. Explicitly,
\begin{equation} \label{L}
L^\lambda{}_{\mu\nu} = \frac{1}{2} \left( Q^\lambda{}_{\mu\nu} - Q_\mu{}^\lambda{}_\nu - Q_\nu{}^\lambda{}_\mu \right) \,.
\end{equation}

Two independent non-metricity vectors can be constructed as
\begin{equation*}
 Q_\mu := g^{\nu\lambda}Q_{\mu\nu\lambda} = Q_\mu{}^\nu{}_\nu \,, \qquad 
 \tilde{Q}_\mu := g^{\nu\lambda}Q_{\nu\mu\lambda} = Q_{\nu\mu}{}^\nu \,,
\end{equation*}
with corresponding contractions of the disformation tensor
\begin{align}
 L_\mu := L_\mu{}^\nu{}_\nu \,, \qquad 
 \tilde{L}_\mu := L_{\nu\mu}{}^\nu \,.   
\end{align}

The superpotential (or non-metricity conjugate) tensor $P^\lambda{}_{\mu\nu}$ is defined as
\begin{equation} \label{P}
P^\lambda{}_{\mu\nu} = \frac{1}{4} \left( -2 L^\lambda{}_{\mu\nu} + Q^\lambda g_{\mu\nu} - \tilde{Q}^\lambda g_{\mu\nu} -\delta^\lambda{}_{(\mu} Q_{\nu)} \right) \,,
\end{equation}
which allows one to construct the non-metricity scalar
\begin{equation} \label{Q}
Q=Q_{\alpha\beta\gamma}P^{\alpha\beta\gamma}\,.
\end{equation}

The above geometrical quantities satisfy the following identities:
\begin{align}
\m R_{\mu\nu}+\m\nabla_\alpha L^\alpha{}_{\mu\nu}-\m\nabla_\nu\tilde L_\mu
+\tilde L_\alpha L^\alpha{}_{\mu\nu}-L_{\alpha\beta\nu}L^{\beta\alpha}{}_\mu=0\,,
\label{mRicci}\\
\m R+\m\nabla_\alpha(L^\alpha-\tilde L^\alpha)-Q=0\,. \label{mR}
\end{align}
Since $Q^\alpha-\tilde Q^\alpha=L^\alpha-\tilde L^\alpha$, one can define the boundary term
\begin{align}\label{aaboundary}
C=\m{R}-Q&=-\m\nabla_\alpha(Q^\alpha-\tilde Q^\alpha) \notag \\
&=-\frac1{\sqrt{-g}}\partial_\alpha\left[\sqrt{-g}(Q^\alpha-\tilde Q^\alpha)\right].
\end{align}
This boundary term provides the link between the Ricci scalar $\m R$ of General Relativity and the non-metricity scalar $Q$.

Motivated by these relations, the $f(Q,C)$ theory extends both $f(Q)$ and $f(\m R)$ gravities into a single unified framework. Its action is given by
\begin{equation}
S=\int \left[ \frac{1}{2\kappa }f(Q,C)+\mathcal{L}_{{\sf M}}\right] \sqrt{-g}\,d^{4}x\,,
\label{eqn:action-fQC}
\end{equation}
where $f$ is an arbitrary function of $Q$ and $C$, and $\mathcal L_{{\sf M}}$ denotes the matter Lagrangian. Variation with respect to the metric yields the field equations
\begin{align}\label{eqn:FE1}
\kappa T_{\mu\nu}
=&-\frac f2g_{\mu\nu}+2P^\lambda{}_{\mu\nu}\nabla_\lambda(f_Q-f_C)
  +\left(\m G_{\mu\nu}+\frac Q2g_{\mu\nu}\right)f_Q \notag\\
  &+\left(\frac C2g_{\mu\nu}-\m\nabla_{\mu}\m\nabla_{\nu}
  +g_{\mu\nu}\m\nabla^\alpha\m\nabla_\alpha \right)f_C\,.
\end{align}
Since the affine connection is independent of the metric tensor in this framework, variation of the action with respect to the connection leads to the second set of field equations,
\begin{align}\label{eqn:FE2-invar}
(\nabla_\mu-\tilde L_\mu)(\nabla_\nu-\tilde L_\nu)
\left[4(f_Q-f_C)P^{\mu\nu}{}_\lambda\right]=0\,,
\end{align}
in the absence of or on the assumption of the conservation of hypermomentum.

In what follows, we aim to investigate the cosmological implications of $f(Q,C)$ gravity. For this purpose, we adopt the spatially flat Friedmann–Robertson–Walker (FLRW) geometry,
\begin{equation}
ds^2=-dt^2+a^2(t)\left[dr^2+r^2(d\theta^2+\sin^2\theta d\phi^2)\right],  \label{3a}
\end{equation}
corresponding to a homogeneous and isotropic matter distribution. Here $a(t)$ is the cosmic scale factor, and its time derivative defines the Hubble parameter $H(t)=\dot{a}(t)/a(t)$, with $\dot{()}$ denoting differentiation with respect to cosmic time $t$.

The FLRW metric (\ref{3a}) exhibits cosmic symmetry, specifically homogeneity and isotropy, which can be described by spatial rotational and translational transformations. A symmetric teleparallel affine connection $\Gamma$ is an affine connection that is both torsion-free and curvature-free. It also possesses both spherical and translational symmetries, implying vanishing of the Lie derivative of the affine connection in terms of the Killing vectors. There exist three classes of such affine connections characterized by a dynamic degree of freedom $\gamma(t)$, as outlined in the references \cite{FLRW/connection, ad/bianchi}
\begin{align} 
\Gamma^t{}_{tt}=&C_1, 
	\quad 					\Gamma^t{}_{rr}=C_2, 
	\quad 					\Gamma^t{}_{\theta\theta}=C_2r^2, 
	\quad						\Gamma^t{}_{\phi\phi}=C_2r^2\sin^2\theta,								\notag\\
\Gamma^r{}_{tr}=&C_3, 
	\quad  	\Gamma^r{}_{rr}=0, 
	\quad		\Gamma^r{}_{\theta\theta}=-r, 
	\quad		\Gamma^r{}_{\phi\phi}=-r\sin^2\theta,												\notag\\
\Gamma^\theta{}_{t\theta}=&C_3, 
	\quad		\Gamma^\theta{}_{r\theta}=\frac1r,
	\quad		\Gamma^\theta{}_{\phi\phi}=-\cos\theta\sin\theta,										\notag\\
\Gamma^\phi{}_{t\phi}=&C_3, 
	\quad 	\Gamma^\phi{}_{r\phi}=\frac1r, 
	\quad 	\Gamma^\phi{}_{\theta\phi}=\cot\theta.
\end{align}

\begin{itemize}
\item $C_1=\gamma,~~C_2=C_3=0$,\label{casesi}\\
\item $C_1=\gamma+\dfrac{\dot\gamma}\gamma,~~C_2=0~~C_3=\gamma,~~~\gamma\neq0$ \label{casesii}\\
\item $C_1=-\dfrac{\dot\gamma}{\gamma},~~C_2=\gamma,~~C_3=0~~~\gamma\neq0$.\label{casesiii}
\end{itemize} 

Let us denote these classes by Connection type $\Gamma_I$, $\Gamma_{II}$ and $\Gamma_{III}$, respectively. For specific cosmological applications, it is necessary to analyze each class of these symmetric teleparallel connections and establish the associated equations of motion for $f(Q,C)$ gravity, which were derived in \cite{als/fQC}. 

Pure quadratic gravity, and in particular pure $R^2$ gravity, continues to provide a compelling arena for exploring gravitational dynamics beyond the Einstein-Hilbert paradigm. Owing to its classical scale invariance and, unlike more general quadratic theories, its ghost-free character, pure $R^2$ gravity possesses several attractive theoretical features and may exhibit a nontrivial interplay between ultraviolet and infrared gravitational physics~\cite{alvarez2016}. Moreover, its connection with scale symmetry breaking, dimensional transmutation, and the possible dynamical emergence of gravitational scales further motivates the investigation of purely quadratic gravitational actions~\cite{einhorn2015}. Recent studies have also shown that, when formulated around an appropriate de Sitter background, pure $R^2$ gravity can support a massless spin-2 mode and yield a Newtonian $1/r$ potential, highlighting the continuing relevance and nontrivial phenomenology of purely quadratic gravity theories~\cite{nguyen2023}. 

Motivated by these considerations, in the following sections we investigate the latter two connection functions by adopting the following quadratic functional form\footnote{Although we do not investigate the field-theoretic aspects of this branch of quadratic $f(Q,C)$ gravity here, the present analysis may offer a useful framework for further developments in this direction.}

\begin{align}\label{Lagrangian}
f(Q,C)=\eta_{1} Q^2+\eta_{2} C^2+\eta_{3} CQ,
\end{align}

\noindent and $\gamma=s(a)$\footnote{Note that the coupling constants $\eta_i$  are dimensionful with $[\eta_i]=[H]^{-2}$, as required by the dimensional consistency of the gravitational action, where ``$[]$'' denotes the dimension of a quantity.},\footnote{For a discussion of the relation between our model and pure $R^2$ gravity, see the second point under the ``main picture'' subsection of the conclusion.}.

\section{Connection class $\Gamma_{II}$}\label{type2}
The equations of motion ({\sf EoM}) corresponding to this class of affine connections take the form~\cite{als/fQC}
 
\begin{align}
&\kappa \rho_{{\sf m}}=\frac{f}{2}+f_Q \left(3 H^2-\frac{Q}{2}\right)-\frac{C f_C}{2}+\frac{3 \gamma  \dot{f}_Q}{2}+\left(3 H-\frac{3 \gamma}{2}\right)\dot{f}_C,\label{t2.1}\\
&\kappa p_{{\sf m}}=-\frac{f}{2}+f_Q \left(-3 H^2-2 H'+\frac{Q}{2}\right)+\frac{C f_C}{2}+\frac{1}{2} (3 \gamma -4 H) \dot{f}_Q-\frac{3 \gamma  \dot{f}_C}{2}-\ddot{f}_C,\label{t2.2}\\
&\kappa\m\nabla_{\mu}T^{\mu}_{t}=\mathcal{C}_{II}, \text{ where}\label{t2.3}\\
&\mathcal{C}_{II}=\frac{3}{2}\gamma\Big[3H\dot{f}_Q+\ddot{f}_Q-3H\dot{f}_C-\ddot{f}_C\Big].\nonumber
\end{align}

For the metric~(\ref{3a}) one obtains

\begin{align}
&Q=-6 H^2+9 \gamma  H+3 \dot{\gamma},\label{t2.4}\\
&C=6 \left(3 H^2+\dot{H}\right)-9 \gamma  H-3 \dot{\gamma},\label{t2.5}\\
&Q+C=\mathring R=6(2H^{2}+\dot{H})\label{t2.6}.
\end{align}

For the choice~(\ref{Lagrangian}) we introduce the following dimensionless variables

\begin{align}\label{t2.7}
&x_1=\frac{Q}{H^2},~~~x_2=\frac{C}{H^2},~~~x_3=\frac{\dot{C}}{H^3},\nonumber\\
&x_4=\frac{\dot{Q}}{H^3},~~~x_5= \frac{\dot{H}}{H^2},~~~x_6=\frac{\gamma}{H},~~~x_7=\frac{\ddot{C}}{H^4}.
\end{align}

and

\begin{align}\label{t2.7.1}
\widetilde{\Omega}_{m}=\frac{\kappa\rho_{m}}{3H^4}.
\end{align}

We now derive the dynamical system by following the steps outlined below.

\subsection{Friedmann equations in terms of dimensionless variables}\label{subsec2.1}
First, substituting the definitions of the dimensionless variables given in~(\ref{t2.7}), and~(\ref{t2.7.1}) into eqs.~(\ref{t2.1})-(\ref{t2.3}), respectively, yields

\begin{align}
&\widetilde{\Omega}_{{\sf m}}=\label{t2.8}\\
& \eta _1\Big(-\frac{x_1^2}{6}+2 x_1+x_4 x_6\Big)+\eta _2\Big(-\frac{x_2^2}{6}-x_3 \left(x_6-2\right)\Big)+\eta _3\Big(-\frac{1}{6} \left(x_1-6\right) x_2+x_4+\frac{1}{2} \left(x_3-x_4\right) x_6\Big),\nonumber\\
&x_{7}=\frac{1}{4 \eta _2}\Bigg\{\bigg[\eta _1\Big(x_1 \left(x_1-8 x_5-12\right)+2 x_4 \left(3 x_6-4\right)\Big)+\eta_{2}\Big(x_2^2-6 x_3 x_6\Big)\nonumber\\
&+\eta_{3}\Big(x_2 \left(x_1-4 x_5-6\right)-3 x_4 x_6+x_3 \left(3 x_6-4\right)\Big)\bigg]-6 w\bigg[\eta_{1}\Big(-\frac{x_1^2}{6}+2 x_1+x_4 x_6\Big)\nonumber\\
&+\eta_{2}\Big(-\frac{x_2^2}{6}-x_3 \left(x_6-2\right)\Big)+\eta_{3}\Big(-\frac{1}{6} \left(x_1-6\right) x_2+x_4+\frac{1}{2} \left(x_3-x_4\right) x_6\Big)\bigg]\Bigg\}-\frac{\eta_3}{2\eta_2}\frac{\ddot{Q}}{H^4},\label{t2.9}\\
&\mathcal{C}_{II}=3\Bigg[\eta_{1} \Big(3x_4 H^4 +\ddot{Q}\Big)-\eta_{2}\Big( 3 x_3+x_7\Big)H^{4}+\frac{\eta_{3}}{2}\Big(\left(3 x_3-3 x_4+x_7\right)H^4 -\ddot{Q}\Big)\Bigg]x_6H,\label{t2.10}
\end{align}

where we have used the definition of the {\sf EoS} parameter, $w=p_{m}/\rho_{m}$. In eq.~(\ref{t2.8}), the subscript ``${\sf m}$" denotes the matter component of the Universe. From the definitions in~(\ref{t2.7}), it follows that deriving the first-order autonomous system in terms of the independent variables requires the evaluation of higher-order time derivatives, such as $\ddot{H}$ and $\ddot{Q}$. These quantities are derived in the following subsections. It is worth noting that $\ddot{Q}$ also appears in the expression for $\mathcal{C}_{II}$ given in~(\ref{t2.10}). To this end, we first note from eq.~(\ref{t2.9}) that $x_7$ is algebraically determined by the remaining variables and is therefore not an independent dynamical variable.\\

\subsection{Constraint equations}\label{subsec2.2}

Differentiating eq.~(\ref{t2.6}) with respect to time and rewriting the resulting expression in terms of the variables defined in~(\ref{t2.7}) yields

\begin{align}\label{t2.11}
\ddot{H}=\frac{1}{6} H^3 \big(x_3+x_4-24 x_5\big).
\end{align}

Then, equating the expressions for $\dot{\gamma}$ derived from eq.~(\ref{t2.4}) and eq.~(\ref{t2.5}) in terms of the variables defined in eq.~(\ref{t2.7}) gives the following constraint:

\begin{align}\label{t2.12}
x_{1}+x_{2}=6 (x_{5}+2).
\end{align}

The second time derivative of $\gamma$ is obtained by differentiating either eq.~(\ref{t2.4}) or eq.~(\ref{t2.5}) twice with respect to time. After applying the equations given in eqs.~(\ref{t2.11}) and (\ref{t2.12}), the two procedures yield an identical relation, expressed as

\begin{align}\label{t2.13}
\ddot{\gamma}=\frac{1}{3} \Big[-3 x_1+x_4+12 x_5-9 \left(x_5-3\right) x_6-18\Big]H^3 ,
\end{align}

\noindent where either Eq.~(\ref{t2.4}) or Eq.~(\ref{t2.5}) has been employed again after differentiation. To obtain the third time derivative of $\gamma$, we differentiate these relations once more. Using Eq.~(\ref{t2.4}), we obtain

\begin{align}\label{t2.14}
\dddot{\gamma}=\left(4 H^3 x_5^2+\left(4-3 x_6\right) \ddot{H}-6 H x_5 \dot{\gamma}-3 \ddot{\gamma}\right)H +\frac{\ddot{Q}}{3}
\end{align}

which contains $\ddot{Q}$. Similarly, Eq.~(\ref{t2.5}) provides an alternative expression, given by

\begin{align}\label{t2.15}
&\dddot{\gamma}=\left(12  x_5^2-\frac{ x_7}{3}\right)H^4-6 H^2 \dot{\gamma} x_5-3 \left(\left(x_6-4\right) \ddot{H}+\ddot{\gamma}\right)H+2 \dddot{H}.
\end{align}

This expression for $\dddot{\gamma}$ contains $\dddot{H}$. Hence, equating Eqs.~(\ref{t2.14}) and (\ref{t2.15}) provides a relation between $\ddot{Q}$ and $\dddot{H}$.
\subsection{Constraints associated with the $\gamma$-function}\label{subsec2.3}

A natural ansatz for the cosmological evolution of $\gamma$ is to assume that it is a function of the scale factor, i.e., $\gamma=s(a)$. Using the relations

\begin{align}\label{t2.16}
&\dot{a}=aH,\\
&\ddot{a}=a(H^{2}+\dot{H}),\\
&\dddot{a}=a\ddot{H}+\frac{\dot{a}}{a^{2}}\left(-2\dot{a}^{2}+3a\ddot{a}\right),
\end{align}

and introducing the following dimensionless variables

\begin{align}\label{t2.17}
&n=\frac{\gamma'a}{\gamma},\nonumber\\
&m=\frac{\gamma''a}{\gamma'},\nonumber\\
&l=\frac{\gamma'''a}{\gamma''},
\end{align}

which provide a parameterization of the $\gamma$-function,  one obtains

\begin{align}\label{t2.18}
\dot{\gamma}=n x_6 H^2.
\end{align}

In definitions~(\ref{t2.17}), primes denote differentiations with respect to the scale factor. Combining the expressions for $\dot{\gamma}$ obtained from eq.~(\ref{t2.18}) and either eq.~(\ref{t2.4}) or eq.~(\ref{t2.5}) gives

\begin{align}\label{t2.19}
6+x_1=3 \left(n+3\right)x_6,
\end{align}

\noindent which provides another constraint whose form depends on the specific functional form of the $\gamma$-function. The expressions for the second- and third-order time derivatives are then obtained as

\begin{align}
&\ddot{\gamma}= n \left(1+m+x_5\right)x_6H^3, \label{t2.20}\\
&\dddot{\gamma}=\frac{n}{6}  \Big[6 m \left(l+3 x_5+3\right)+x_3+x_4-6 x_5+6\Big]x_6H^4. \label{t2.21}
\end{align}

By equating eq.~(\ref{t2.20}) to eq.~(\ref{t2.13}), we obtain

\begin{align}\label{t2.22}
x_4+12 x_5=3  \Big[n \left(4+m+x_5\right)+3 x_5\Big]x_6.
\end{align}

Similarly, equating the expressions for $\dddot{\gamma}$ obtained from eqs.~(\ref{t2.14}) and~(\ref{t2.21}) yields

\begin{align}\label{t2.23}
&\ddot{Q}=\frac{1}{2}\Bigg\{x_6 \bigg[n \Big(6 m \left(6+l+3 x_5\right)+x_3+x_4+48 x_5+24\Big)+3 \left(x_3+x_4-24 x_5\right)\bigg]\\
&-4 \Big(x_3+x_4+6 \left(x_5-4\right) x_5\Big)\Bigg\}H^4.\nonumber
\end{align}

\subsection{Final dynamical equations}\label{subsec2.4}

From the above discussion, we conclude that the dimensionless variables are not all independent. Due to the constraint equations~(\ref{t2.8}),~(\ref{t2.9}),~(\ref{t2.12}),~(\ref{t2.19}) and~(\ref{t2.22}), only three independent variables remain out of the eight variables introduced in eq~(\ref{t2.7}). We are therefore free to choose any three variables as independent variables. Here, we select $x_3$, $x_5$ and $x_6$ since the latter two characterize the evolution of the {\sf EoS} parameter and the $\gamma$-function, respectively. The choice of $x_3$ is made for convenience. Finally, by adopting $w=0$ , the resulting dynamical system takes the following closed form

\begin{align}
&\frac{dx_{3}}{dN}=\frac{1}{4 \eta_2}\left(\eta_1\mathcal{A}_1+\eta_2\mathcal{A}_2+\eta_3\mathcal{A}_3\right),\label{t2.24}\\
&\mathcal{A}_1=3 \Bigg\{48 x_5+36 -8 \Big[(m+7) n+(2 n+9) x_5+9\Big]x_6+3\Big[2 (m+7) n+n^2+2 (n+3) x_5+9\Big] x_6^2 \Bigg\},\nonumber\\
&\mathcal{A}_2=3 \Bigg\{12\Big[x_5 \left(3 \left(x_5+6\right)-x_3\right)+27\Big] -6 \Big[6 (n+3) \left(x_5+3\right)+x_3\Big]x_6 +9 (n+3)^2 x_6^2\Bigg\},\nonumber\\
&\mathcal{A}_3=3216+\Bigg\{19 n x_3-270+6 n \Big[(l+4) m-19\Big]+3 \Big[(m+7) n (n+6)+27\Big]x_6 \Bigg\}x_6+ \nonumber\\
&3 \Bigg\{9+\Big[6 m n+(n+3) (n+6) x_6-8 n-90\Big]x_6 \Bigg\}x_5, \nonumber\\
&\frac{dx_{5}}{dN}=\frac{1}{6} \Big\{3 x_6 \big[(m+4) n+(n+3) x_5\big]+x_3-12 x_5 \left(x_5+3\right)\Bigg\},\label{t2.25}\\
&\frac{dx_{6}}{dN}=(n-x_5)x_6.\label{t2.26}
\end{align}

In the following subsection, we analyze the critical points of the dynamical system given by eqs.~(\ref{t2.24})-(\ref{t2.26}).
%
\subsection{Critical points}\label{subsec2.5}
Equation~(\ref{t2.26}) admits two branches of solutions, namely $x_5=n$ as well as $x_6=0$. For the first branch, one obtains $w_{eff}=-1-2\dot{H}/3H^2=-1-2n/3$. This relation indicates that the system admits a pressureless-fluid solution, corresponding to a {\sf DM}-dominated era, for $n=-3/2$. Therefore, to ensure that the system admits a dust-like critical point, we fix the functional form of $\gamma$ by choosing $n=-3/2$. Consequently $\gamma=\gamma_{0}a^{-3/2}$ which gives $m=-5/2$ and $l=-7/2$. As a result, the complete critical-point structure is summarized in Table~\ref{tab1}.

\begin{table}[h!]
\centering
\caption{Fixed-point solutions of the dynamical system described by eqs.~(\ref{t2.24})-(\ref{t2.26}) for $n=-3/2$, $m=-5/2$, and $l=-7/2$.}
\begin{adjustbox}{width=1\textwidth}
\begin{tabular}{l @{\hskip 0.1in}l @{\hskip 0.1in} l @{\hskip 0.1in}l @{\hskip 0.1in}@{\hskip 0.1in}ll}\hline\hline

Fixed point     &$(x_3,x_5,x_6)$  &$\widetilde{\Omega}_{\sf m}$      &$w_{eff}$\\[0.5 ex]
\hline
$P_1$       & $\left(-\frac{81}{4},-\frac{3}{4},0\right)$ 
                     & $9 \left(-2 \eta _1+4 \eta _3-\frac{63 \eta _2}{8}\right)$
            & $-\frac{1}{2}$\\[0.75 ex]
$P_2$    & $\left(12 \frac{-\mathcal{\widetilde {B}}_1+\eta _1+2 \eta _2-2 \eta _3}{\eta _2},-\frac{\mathcal{B}_1}{\eta _2}-1,0\right)$        
                        & $0$ 
            & $\frac{2 \mathcal{\widetilde {B}}_1}{3 \eta _2}-\frac{1}{3}$\\[0.75 ex]
$P_3$    &  $\left(12 \frac{\mathcal{\widetilde {B}}_1+\eta _1+2 \eta _2-2 \eta _3}{\eta _2},\frac{\mathcal{\widetilde {B}}_1}{\eta _2}-1,0\right)$          
                        & $0 $
            & $-\frac{2 \mathcal{\widetilde {B}}_1}{3 \eta _2}-\frac{1}{3}$\\[0.75 ex]
$P_4$    & $\left(\frac{9}{2}\frac{\mathcal{\widetilde {B}}_2-2 \eta _1+2 \eta _2+19 \eta _3}{\eta _1+\eta _2-7 \eta _3},-\frac{3}{2},\frac{1}{3}\frac{\mathcal{\widetilde {B}}_2+4 \eta _1+8 \eta _2-23 \eta _3}{\eta _1+\eta _2-7 \eta _3}\right)$              
                        & $\widetilde{\Omega}_4$
            & $0$ \\[0.75 ex]
$P_5$    & $\left(\frac{9}{2}\frac{-\mathcal{\widetilde {B}}_2-2 \eta _1+2 \eta _2+19 \eta _3}{\eta _1+\eta _2-7 \eta _3},-\frac{3}{2},\frac{1}{3}\frac{-\mathcal{\widetilde {B}}_2+4 \eta _1+8 \eta _2-23 \eta _3}{\eta _1+\eta _2-7 \eta _3}\right)$              
                        & $\widetilde{\Omega}_5$& $0$  \\[0.75 ex]
\hline\hline
\end{tabular}
\end{adjustbox}
\label{tab1}
\end{table}

In Table~\ref{tab1} we have defined

\begin{align}
&\mathcal{\widetilde {B}}_1=\sqrt{\eta _2 \left(\eta _1+4 \eta _2-2 \eta _3\right)},\label{t2.27}\\
&\mathcal{\widetilde {B}}_2=\sqrt{4 \eta _2^2-12 \eta _1 \eta _2+84 \eta _3 \eta _2+305 \eta _3^2-40 \eta _1 \eta _3},\label{t2.28}\\
&\widetilde{\Omega}_4=\frac{3}{2 \left(\eta _1+\eta _2-7 \eta _3\right)^2}\Bigg\{\left(\eta _1+\eta _2\right) \bigg[-\mathcal{B}_2 \left(2 \eta _1+5 \eta _2\right)+2 \eta _2 \left(\eta _1-5 \eta _2\right)\bigg]\nonumber\\
&\bigg[\eta _2 \left(-39 \mathcal{B}_2+11 \eta _1\right)+38 \eta _1^2-183 \eta _2^2\bigg]\eta _3 -\left(-23 \mathcal{B}_2+372 \eta _1+615 \eta _2\right)\eta _3^2 +367 \eta _3^3\Bigg\},\label{t2.29}\\
&\widetilde{\Omega}_5=\frac{3}{2 \left(\eta _1+\eta _2-7 \eta _3\right)^2}\Bigg\{\left(\eta _1+\eta _2\right) \bigg[\mathcal{B}_2 \left(2 \eta _1+5 \eta _2\right)+2 \eta _2 \left(\eta _1-5 \eta _2\right)\bigg]\nonumber\\
&\bigg[\eta _2 \left(39 \mathcal{B}_2+11 \eta _1\right)+38 \eta _1^2-183 \eta _2^2\bigg]\eta _3 -\left(23 \mathcal{B}_2+372 \eta _1+615 \eta _2\right)\eta _3^2 +367 \eta _3^3\Bigg\}.\label{t2.30}
\end{align}


\subsection{Fixed-point analysis, phase-space connections and $\epsilon$-parameterization}\label{subsec2.6}

As shown in Table~\ref{tab1}, the model admits several possible cosmological evolutions connecting a {\sf DM}-dominated phase, characterized by $w_{eff}=0$ and $\widetilde{\Omega}_{{\sf m}}>0$, to an accelerated-expansion ({\sf AE}) phase, characterized by $w_{eff}<-1/3$ and $\widetilde{\Omega}_{{\sf m}}=0$. In particular, phase-space trajectories may connect any of the fixed points $P_1$, $P_2$ and $P_3$ to either $P_4$ or $P_5$, provided that the corresponding physical requirements are satisfied. For example, for the transition $P_1\to P_4/P_5$, the matter density parameter a $P_1$ must satisfy $\widetilde{\Omega}_{{\sf m}}^{P_1}=0$ which holds only within a specific region of the parameter space of the model, $\eta_i$. Furthermore, for the same region of the parameter space, $P_4$ or $ P_5$ must be unstable or saddle, while $P_1$ should be stable, thereby allowing the corresponding physically meaningful phase-space trajectory. In addition, to ensure the conservation of the {\sf EMT}, the condition $\mathcal{C}_{II}=0$ must be satisfied (see eq.~(\ref{t2.3}) and~(\ref{t2.10})). Since identifying the parameter ranges that satisfy all the aforementioned conditions requires an extensive analysis, and since not all possible connections need to be explored, we restrict our study to a selected case. Therefore, we restrict our analysis to the transition $P_2/P_3\to P_4$. It is worth noting that $w_{eff}^{P_1}=-1/2$, which is incompatible with the recent observational constraints on the {\sf EoS} parameter. Therefore, we exclude the connections from $P_1$ from further consideration. We then rewrite the the {\sf EoS} parameter associated with $P_2/P_3$ by introducing the $\epsilon$-parameter to describe its deviation from $-1$, as follows

\begin{align}
&w_{eff}^{P_2}=-1+\epsilon,~~~~~~~\to~~~~~~~~\eta _3=\frac{1}{8} \Big[4 \eta _1-3 \eta _2 (\epsilon -2) (3 \epsilon +2)\Big]~~~\mbox{for}~~~\eta_2<0~~\mbox{and}~~\epsilon<\frac{2}{3},\label{t2.31.a}\\
&w_{eff}^{P_3}=-1+\epsilon,~~~~~~~\to~~~~~~~~\eta _3=\frac{1}{8} \Big[4 \eta _1-3 \eta _2 (\epsilon -2) (3 \epsilon +2)\Big]~~~\mbox{for}~~~\eta_2>0~~\mbox{and}~~\epsilon<\frac{2}{3}.\label{t2.31.b}
\end{align}

The results presented in eqs.~(\ref{t2.31.a}) and~(\ref{t2.31.b}) show that the fixed points $P_2$ and $P_3$ represent the {\sf AE} phase for opposite signs of the parameter $\eta_2$. Hence, by replacing $\eta _3$ with the corresponding expression in terms of $\epsilon$, we reduce the parameter space such that the {\sf AE} solution is guaranteed for sufficiently small $\epsilon$. As the next step, selecting $P_4$ as the {\sf DM}-dominated solution, we rewrite the expressions for $\widetilde{\Omega}_4$ and $\mathcal{C}_{II}$ evaluated at $P_4$, and eq.~(\ref{t2.24}) in terms of $\epsilon$, which take the following form

\begin{align}\label{t2.31}
&\widetilde{\Omega}_4=\frac{3}{16 \Big[20 \eta _1+\eta _2 \big(21 \epsilon  (4-3 \epsilon )+76\big)\Big]^2}\Bigg\{240 \eta _1^2 \bigg[\eta _2\Big (47 \epsilon  (3 \epsilon -4)-500\Big)+\mathcal{G}\bigg]-14400 \eta _1^3\nonumber\\
&+4 \eta _1 \eta _2 \bigg[\eta _2 \Big(3 \epsilon  (3 \epsilon -4) \big(1071 \epsilon  (3 \epsilon -4)+1096\big)-77296\Big)+2 \mathcal{G} \Big(69 \epsilon  (4-3 \epsilon )+64\Big)\bigg]\nonumber\\
&+\eta _2^2\bigg[\eta _2 \Big(-9 \epsilon  (3 \epsilon -4) \big[3 \epsilon  (3 \epsilon -4) \big(367 \epsilon  (3 \epsilon -4)-2764\big)+9584\big]-219968\Big)\nonumber\\
&+\mathcal{G} \Big(9 \epsilon  (3 \epsilon -4) \big(23 \epsilon  (3 \epsilon -4)-80\big)-752\Big)\bigg]\Bigg\},
\end{align}

\begin{align}\label{t2.32}
&\mathcal{C}_{II}=\frac{\mathfrak{T}_0+\mathfrak{T}_1 \mathcal{G}+\mathfrak{T}_2 \mathcal{G}^2+\mathfrak{T}_3 \mathcal{G}^3}{2048 \eta _2 \Big[\eta _2 \big(21 \epsilon  (3 \epsilon -4)-76\big)-20 \eta _1\Big]^3}H^5,
\end{align}

\begin{align}\label{t2.33}
&\frac{dx_{3}}{dN}=\frac{3}{64}\bigg\{\frac{\eta _1 }{\eta _2}\big[9 x_6 (10 x_5-9)+76 x_3-54 (6 x_5+1)\big]x_6 +64  \big[3 (x_5-6)-x_3\big]x_5\nonumber\\
&+27 \left(7-6 x_5\right) x_6^2+2 \left(98 x_3+522 x_5-81\right) x_6\bigg\}\nonumber\\
&+\frac{9}{64} \Bigg[x_6 \left(76 x_3+27 x_6+234\right)-6 x_5 \bigg(x_6 \left(9 x_6-74\right)+128\bigg)-576\Bigg]\epsilon\nonumber\\
&-\frac{27}{256} \Bigg[x_6 \left(76 x_3+27 x_6+234\right)-6 x_5 \bigg(x_6 \left(9 x_6-74\right)+128\bigg)+576\Bigg]\epsilon^2.
\end{align}

where we have defined

\begin{align}\label{t2.34}
\mathcal{G}=\sqrt{3600 \eta _1^2+120 \eta _2 \eta _1 \big(53 \epsilon  (4-3 \epsilon )+228\big)+\eta _2^2 \Big[9 \epsilon  (3 \epsilon -4) \big(305 \epsilon  (3 \epsilon -4)-2664\big)+52240\Big]}.
\end{align}

The explicit forms of $\mathfrak{T}_i$ appearing in eq.~(\ref{t2.31}) are provided in Appendix~\ref{app1}. It should be noted that $dx_{5}/dN$ and $dx_{6}$ are independent of the parameters $\eta_i$, and therefore of the $\epsilon$-parameter (see eqs.~(\ref{t2.25}) and~(\ref{t2.26})). Another important point is that, according to the expression~(\ref{t2.10}), $\mathcal{C}_{II}=0$ at the critical points $P_1$, $P_2$ and $P_3$, since $x_6=0$ at these points. Therefore, imposing the conservation of the {\sf EMT} at the critical points requires the fixed point associated with the {\sf DM} era also respects the rule. Accordingly, since $P_4$ has already been identified as the representative of the {\sf DM} era, we determine the parameter values $\eta_1$ and $\eta_2$ for given $\epsilon$ , by imposing the conditions $\mathcal{C}_{II}(P_4)=0$ and $\widetilde{\Omega}_4>0$, together with eqs.~(\ref{t2.31}) and~(\ref{t2.32}). To facilitate the subsequent discussion, we further adopt the normalization $\widetilde{\Omega}_4=1$.

\section{Numerical study of the cosmological evolution}\label{numerical}

\begin{figure}[ht!]
\centering
\epsfig{figure=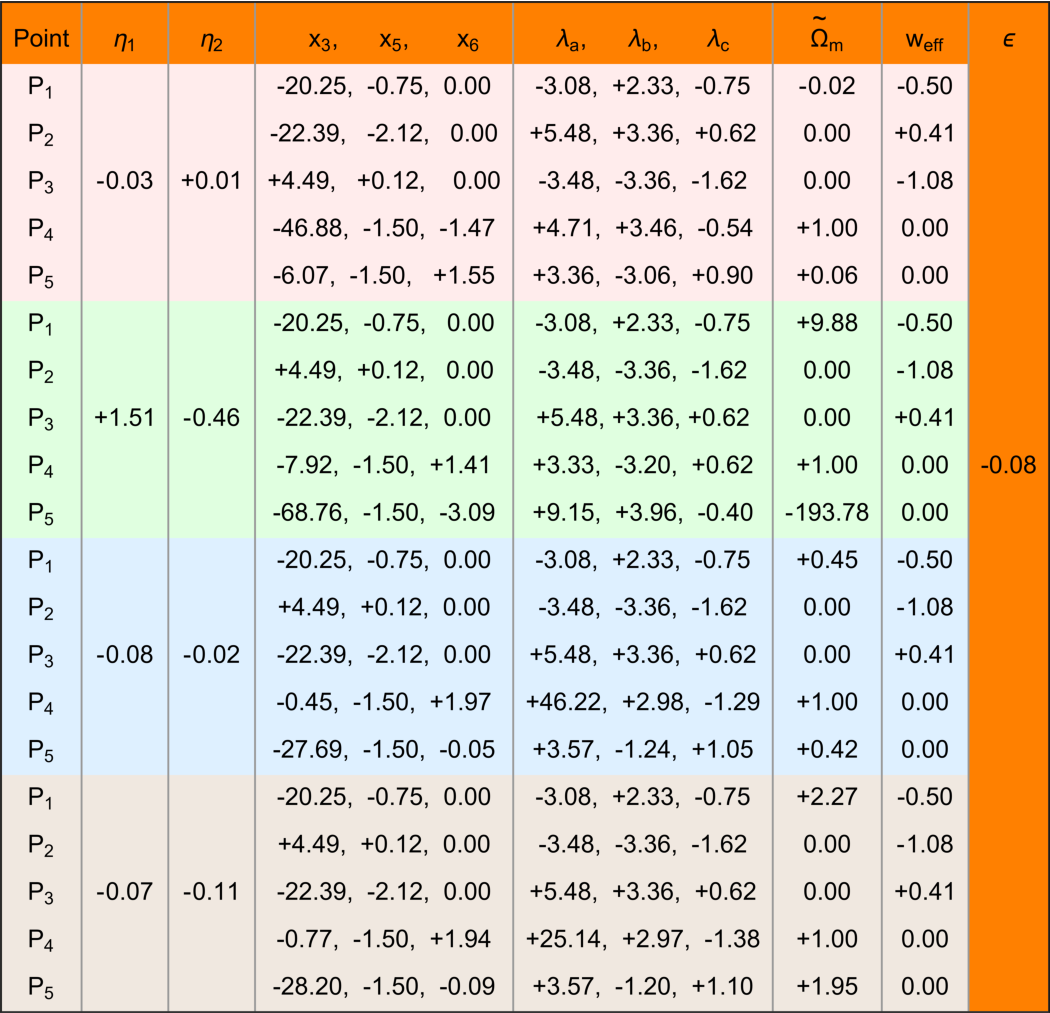,width=12cm}
\caption {Fixed-point solutions of the dynamical system described by eqs.~(\ref{t2.25})-(\ref{t2.26}) and~(\ref{t2.33}) for $n=-3/2$, $m=-5/2$ and $l=-7/2$ for $\epsilon=-0.08$. Different sets of solutions are distinguished by different colors. The fixed points $P_i$ correspond to those listed in Table~\ref{tab1}. The quantities $\lambda_i$ denote the eigenvalues of the Jacobian matrix evaluated at each fixed point.} \label{fig1}
\end{figure}

\begin{figure}[ht!]
\centering
\epsfig{figure=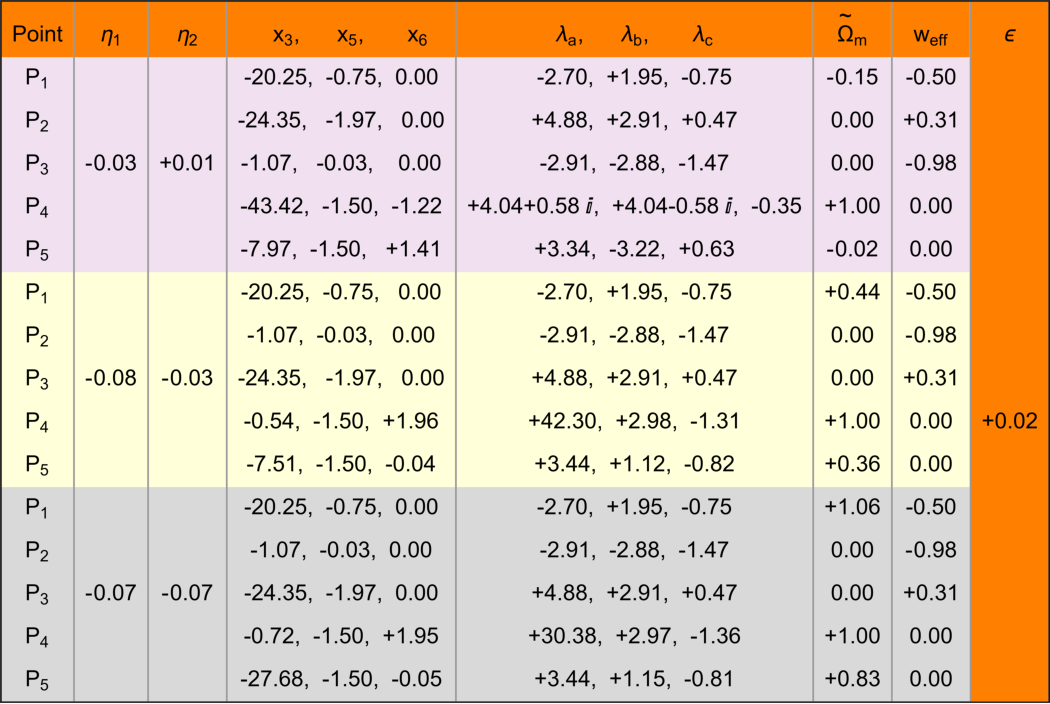,width=13cm}
\caption {Same as Fig.~\ref{fig1}, but for $\epsilon=+0.02$.}\label{fig2}
\end{figure}

Based on the clarifications provided in the final paragraph of the previous subsection, we conclude that, for a given value of $\epsilon$, several sets of $\eta_1$ and $\eta_2$ may exist such that the {\sf EoM}, namely eqs.~(\ref{t2.25})-(\ref{t2.26}) and~(\ref{t2.33}), yield the desired cosmological evolution. In other words, each set of $\eta_1$ and $\eta_2$ values requires a separate analysis of the corresponding {\sf EoM}. For example, the corresponding results for $\epsilon=-0.08$ and $\epsilon=+0.02$ are presented in Figs.~\ref{fig1} and~\ref{fig2}. In these tables, $x_3$, $x_5$ and $x_6$ denote the coordinates of the fixed points, while $\lambda_a$, $\lambda_b$ and $\lambda_c$ represent their corresponding eigenvalues. The different solution sets are distinguished by different colors.\\

Based on the choices in~(\ref{t2.31.a}) and~(\ref{t2.31.b}), the following observations can be extracted from the tables presented in Figs.~\ref{fig1} and~\ref{fig2}.

\begin{itemize}

\item 
The value of $\widetilde{\Omega}_{{\sf m}}$ at $P_1$ depends on the selected ($\eta_1$, $\eta_2$) parameter set. In fact, one obtains
\begin{align}\label{t2.35}
\widetilde{\Omega}_{{\sf m}}^{P_1}=-\frac{27}{8}  (2 \epsilon -1) (6 \epsilon -5)\eta _2 ,
\end{align}
which explicitly depends on the parameter $\eta_2$. For $0<\widetilde{\Omega}_{{\sf m}}^{P_1}<1$, the point $P_1$ corresponds to a state containing both nonzero matter and $w_{eff}=-1/2$. This occurs, for example, in the case $\epsilon=-0.08$ with $\eta_1=-0.08$ and $\eta_2=-0.02$  (the blue region in Fig~\ref{fig1}), as well as in the case $\epsilon=+0.02$ with $\eta_1=-0.08$ and $\eta_2=-0.03$ (the yellow region in Fig~\ref{fig2})).

\item 

The coordinates and eigenvalues $P_2$ and $P_3$ remain unchanged in all cases. In fact, they are independent of the parameters  $\eta_1$ and $\eta_2$. For example, for $P_2$ we obtain
\begin{align}\label{t2.36}
x_3^{P_2}=27 (\epsilon -2) \epsilon,~~~~~x_5^{P_2}=-\frac{3}{2}\epsilon,
\end{align}

for $\eta_2<0$ (see the conditions in~(\ref {t2.31.a})). This fact indicates that $P_2$ and $P_3$ exist for all values of the $\eta_i$ parameters. Therefore, the underlying model always admits an {\sf AE} solution, independently of the specific values of these parameters. We also emphasize that $P_2$ for $\eta_2>0$ and $P_3$ for $\eta_2<0$ may not be assigned a physically meaningful role, since although $\widetilde{\Omega}_{{\sf m}}=0$ at these points, the effective {\sf EoS}
 
\begin{align}\label{t2.37}
w_{eff}^{P_2,~\eta_2>0}=\frac{1}{3}-\epsilon=w_{eff}^{P_3,~\eta_2<0},
\end{align}

remains approximately $1/3$ for small values of $\epsilon$\footnote{Although this case can be interpreted as an {\sf AE} state accompanied by an ultra-relativistic cosmic fluid, such a possibility is not considered in the present work.}. Moreover, Figs.~\ref{fig1} and~\ref{fig2} indicate that, for a given value of $\eta_2$, the fixed point among $P_2$ and $P_3$ satisfying $w_{eff}=-1+\epsilon$ is stable (all three eigenvalues are negative), whereas the other fixed point is a saddle point. This provides another indication of the universal nature of the {\sf EA} solution in the model. For example, when $P_2$ is the {\sf AE} point, corresponding to $\eta_2<0$ (or alternatively $P_3$ for $\eta_2>0$), the eigenvalues are given by

\begin{align}\label{t2.38}
\lambda_{a,b,c}^{P_2,~\eta_2<0}=\frac{3 (\epsilon -1)}{2},~~\frac{9 \epsilon }{2}-3,~~6 \epsilon -3=\lambda_{a,b,c}^{P_3,~\eta_2>0},
\end{align}
which indicates that these points are stable for sufficiently small values of $\epsilon$.

\item 
The fixed points $P_4$ and $P_5$ have different coordinates and eigenvalues for different sets of $\eta_i$ parameters. This indicates that the existence and stability properties of the $\sf DM$ solution depend on the specific parameter values; therefore, unlike the {\sf AE} solution, it is not a universal solution of the model. The {\sf DM} point $P_4$ is an unstable critical point, reflecting the transient nature of the {\sf DM} era. Additionally, $x_6$ at the matter point $P_4$ is nonzero, in contrast to the {\sf AE} points where $x_6=0$. Since $x_6\propto\gamma$ , this indicates that the existence of $P_4$ (and $P_5$) is directly related to the specific assumptions regarding the affine connection and the $\gamma$-function adopted in the $f(Q,C)$ theory.

\item 
Here, we have imposed the conditions under which the point $P_4$ represents a {\sf DM}-dominated era. A similar analysis can be performed by imposing appropriate conditions such that $P_5$ also represents a {\sf DM}-dominated era.

\end{itemize}

In the following subsections, we numerically analyze the cosmological evolution and phase-space behavior of our model.

\subsection{Phase-space diagrams}\label{numerical-1}

A standard approach for analyzing the behavior of phase-space trajectories defined in terms of the original variables (see Eq.~(\ref{t2.7})) is to compactify the phase space by introducing the Poincar\'e coordinates. In practice, one defines a new set of dimensionless variables that maps the boundaries of the original phase space, located at infinity, onto a finite domain. This compactification enables the analysis of the asymptotic behavior of phase-space trajectories. Accordingly, we introduce the following variables

\begin{align}\label{t2.39}
x_3=\frac{X_3}{\sqrt{1-X_1^2-X_2^2-X_3^2}}, ~~~x_5=\frac{X_5}{\sqrt{1-X_1^2-X_2^2-X_3^2}},~~~x_6=\frac{X_6}{\sqrt{1-X_1^2-X_2^2-X_3^2}}.
\end{align}

\noindent Accordingly, one obtains a system of differential equations of the form

\begin{align}\label{t2.39.a}
\frac{d X_{i}}{dN}=f_i(X_3,X_5,X_6),~~~~~~i=3,5,6,
\end{align}

\noindent where the Poincar\'e coordinates satisfy the bounds $-1\le X_i\le1$. The corresponding compactified phase space is therefore mapped onto the interior of the unit sphere, whose boundary represents the points at infinity of the original phase space. This compactification enables a complete analysis of the asymptotic behavior of the phase-space trajectories. The explicit expressions for the functions $f_i(X_3,X_5,X_6)$ are presented in eqs.~(\ref{app2})-(\ref{app4}) of Appendix.

\begin{figure}[h!]
\centering
\includegraphics[width=.45\textwidth]{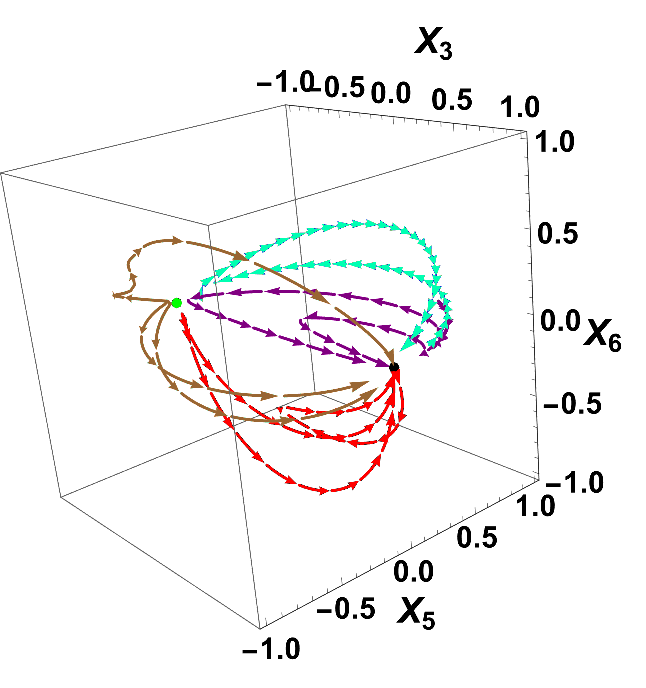}\hspace{2mm}
\includegraphics[width=.45\textwidth]{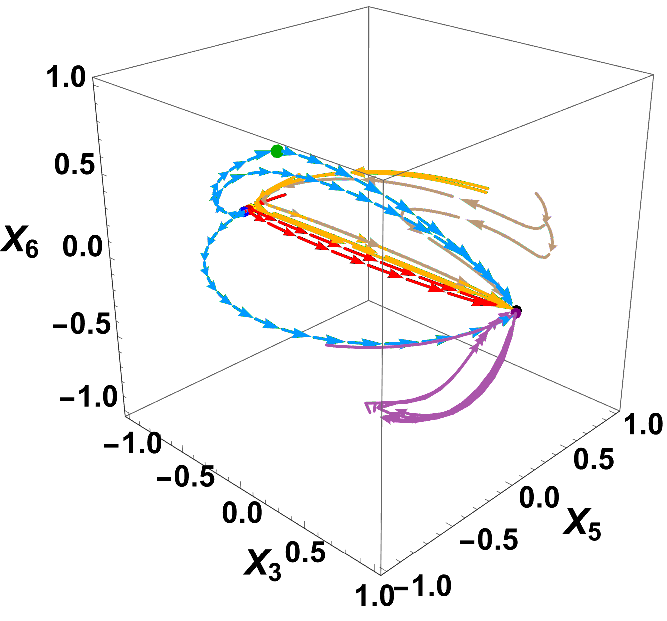}\vspace{2mm}
\includegraphics[width=.45\textwidth]{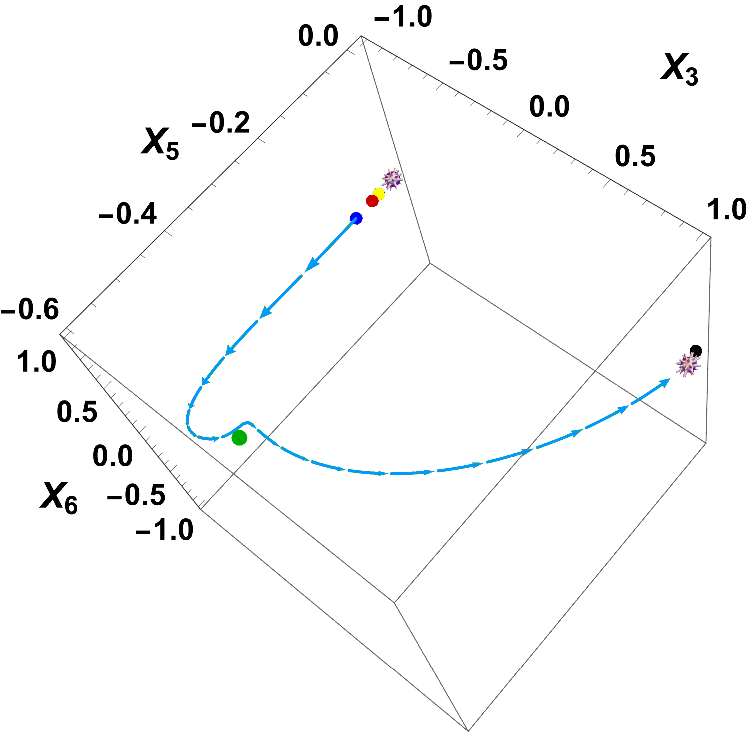}\hspace{2mm}
\includegraphics[width=.45\textwidth]{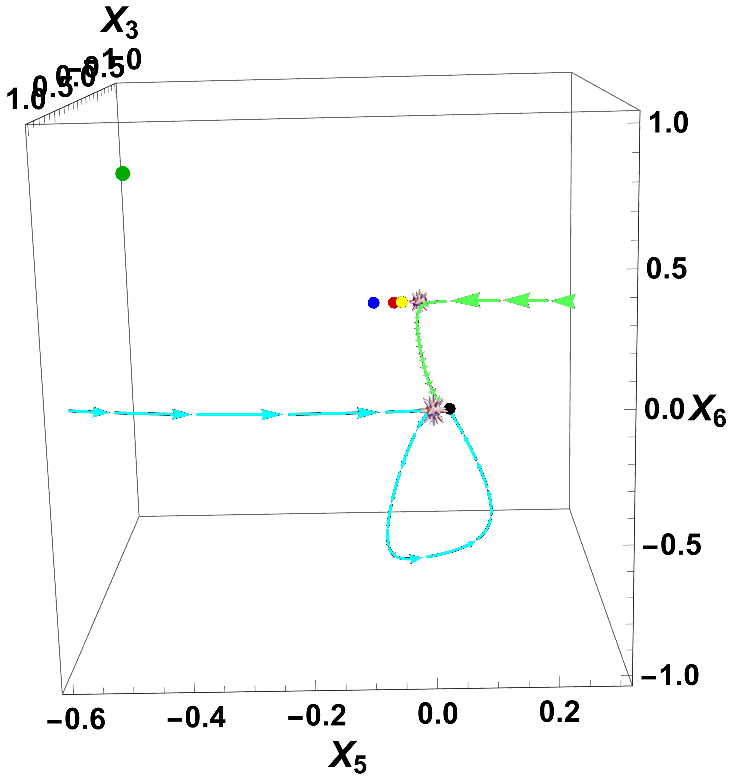}
\caption{Phase-space portraits in the Poincar\'e coordinates defined by eqs.~(\ref{t2.39}) for $\epsilon=-0.08$. The upper-left panel corresponds to $\eta_1=-0.03$ and $\eta_2=+0.01$, whereas the upper-right panel shows the case $\eta_1=-0.08$ and $\eta_2=-0.02$. The lower-left panel is an enlarged view of the upper-right panel, highlighting a particular trajectory. The blue, yellow, red, and black points represent the fixed points $P_3$, $P_1$, $P_5$ and $P_2$, respectively. The light-purple star-shaped markers denote the fixed points located at infinity. The upper-left panel illustrates a $P_4\to P_2$ transition, whereas the upper-right and lower-left panels exhibit a $P_3\to P_4\to P_2$ transition sequence.}\label{fig3}
\end{figure}

Two cases with $\epsilon=-0.08$ and opposite signs of $\eta_2$ are presented in Fig.~\ref{fig3}. The upper-left panel corresponds to $\eta_1=-0.03$ and $\eta_2=+0.01$, whereas the upper-right panel corresponds to $\eta_1=-0.08$ and $\eta_2=-0.02$. As can be observed, the {\sf AE} point, represented by the black marker, acts as the final attractor of all trajectories. The green point, corresponding to the $\sf DM$ solution, evolves toward the {\sf AE} point. In the upper-left panel, the brown trajectories show that the matter era can serve as an initial state, whereas the other trajectories indicate that it may also appear as an intermediate stage. \\

The upper-right panel demonstrates that the matter era is always a transient stage for a class of trajectories, whose initial states correspond to one of the three remaining critical points. As indicated in Fig.~\ref{fig1}, $P_3$ (in the blue region) behaves as an unstable critical point, as all cosmologically relevant trajectories originate from $P_3$, pass through the saddle point $P_4$, and eventually approach $P_2$. The point $P_4$ which is a saddle point can be reached after $P_3$. By magnifying the upper-right panel of Fig.~\ref{fig3} and isolating the trajectory that passes through $P_4$, the lower-left panel explicitly shows the sequence $P_3\to P_4\to P_2$. Here, the blue, yellow and red markers represent $P_3$, $P_1$ and $P_5$, respectively. \\

Furthermore, it is useful to analyze the asymptotic behavior of the phase space at infinity. Following the approach adopted in~\cite{genly2014}, we first transform the equations of motion into spherical coordinates using the following definitions

\begin{align}\label{t2.39.b}
X_3=rsin(\theta)cos(\phi),~~~~X_5=rsin(\theta)sin(\phi),~~~~X_6=rcos(\theta).
\end{align}

Next, the radial coordinate $r$ is replaced by the normalized compactified coordinate defined by

\begin{align}\label{t2.40}
\rho=\frac{r}{\sqrt{1+r^2}}.
\end{align}

\noindent By keeping only the leading-order terms in the limit $\rho\to1$, we obtain the equations governing the dynamics on the boundary of the compactified phase space. The resulting equations are given by

\begin{align}
&\frac{d\rho}{d N}\approx\frac{3 }{2}\cos ^2(\theta )+\frac{1}{12}  \Big[\big(162 \epsilon  (4-3 \epsilon )+323\big) \sin (2 \phi )-36 \cos (2 \phi )+36\Big]\sin ^2(\theta )\label{t2.40.a}\\
&+\frac{9}{1024\eta_2 }\bigg[64 \eta_2 \sin (\phi )+36 \Big(4 \eta_1+\eta_2 \big(13 \epsilon  (3 \epsilon -4)+12\big)\Big)\cos (\phi )+3 \eta_2 \mathcal{K}\sin (2 \theta ) \sin (2 \phi )\bigg]\sin (2 \theta ),\nonumber\\
&\frac{d\theta}{d N}\approx\frac{27 \mathcal{K}}{256\left(1-\rho ^2\right)}\cos ^3(\theta ) \sin (\theta ) \sin (2 \phi ),\label{t2.40.b} \\
&\frac{d\phi}{d N}\approx-\frac{27  \mathcal{K}}{128  \left(1-\rho ^2\right)}\cos ^2(\theta ) \sin ^2(\phi ),\label{t2.40.c}
\end{align}

where $\mathcal{K}= \left[20 \eta_1+9 \eta_2 (\epsilon -2) (3 \epsilon +2)\right]/\eta_2$. As can be seen, only the last two equations contain divergent terms, whereas the first equation depends solely on $\theta$ and $\phi$. Therefore, following Ref.~\cite{genly2014} , we restrict our analysis to the angular subsystem. The stability of the asymptotic solutions is then determined by analyzing the singular points of this two-dimensional system and by examining the sign of eq.~(\ref{t2.40.a}), namely $d\rho/dN$, at the corresponding fixed points.

The system of eqs.~(\ref{t2.40.b})-(\ref{t2.40.c}) admits two solutions with coordinates $(\theta,\Phi)=(\frac{\pi }{2},0)$ and $(\frac{\pi }{2},\pi)$ cooresponding to $(X_3,X_5,X_6)=(1,0,0)$ and $(X_3,X_5,X_6)=(-1,0,0)$, respectively, within the standard intervals\footnote{Note that these points are distinct from the poles of the spherical coordinate system, located at $(X_3,X_5,X_6)=(0,0,\pm1)$.} $0\leq\theta\leq\pi$ and $0\leq\phi<2\pi$. The eigenvalues of the reduced system are $(0,0)$ for both solutions, while $d\rho/dN=0$ at both points. Therefore, the stability of these solutions cannot be determined by this approach. Moreover, the effective {\sf EoS} parameter corresponding to these points is given by

\begin{align}\label{t2.41}
w_{eff}=-1-\frac{2}{3}x_5=-1-\frac{2 \rho  \sin (\theta ) \sin (\phi )}{3 \sqrt{1-\rho ^2}}.
\end{align}

\noindent In the limit $\rho\to1$, the effective {\sf EoS} parameter approaches $-1$ for both $(\theta,\Phi)=(\frac{\pi }{2},0)$ and $(\frac{\pi }{2},\pi)$. Therefore, the fixed points at infinity exhibit the characteristics of cosmological states with $w_{eff}=-1$. These points are represented by the light-purple star-shaped markers in the lower panels of Fig.~\ref{fig3}. Furthermore, the lower-right panel which is plotted for $\epsilon=-0.08$ $\eta_1=-0.08$ and $\eta_2=-0.02$ shows that these points are unstable. Depending on the initial conditions, the trajectories either terminate at $P_2$ directly or first approach one of the fixed points at infinity before eventually converging to $P_2$.

\subsection{Cosmological quantities}\label{numerical-2}

\begin{figure}[h!]
\centering
\includegraphics[width=.45\textwidth]{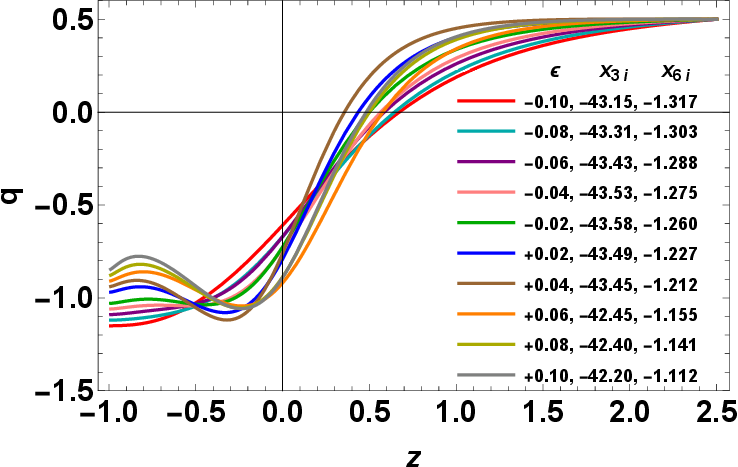}\hspace{2mm}
\includegraphics[width=.45\textwidth]{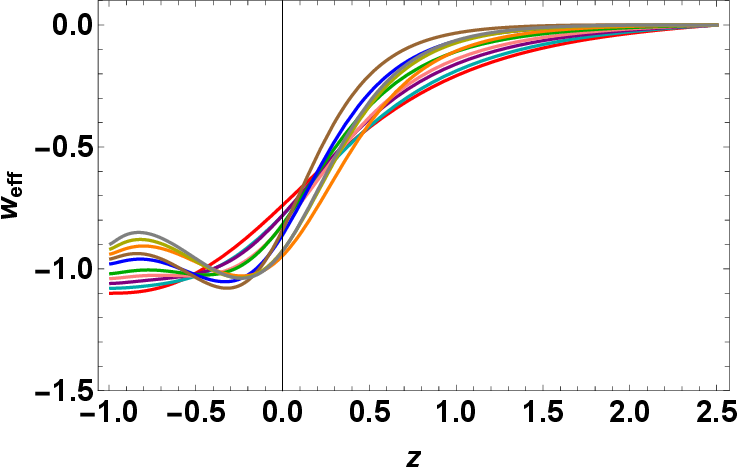}\vspace{2mm}
\includegraphics[width=.45\textwidth]{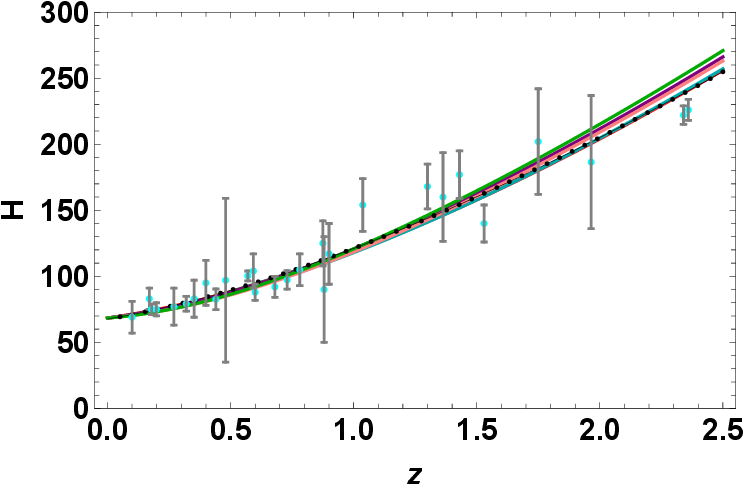}\hspace{2mm}
\includegraphics[width=.45\textwidth]{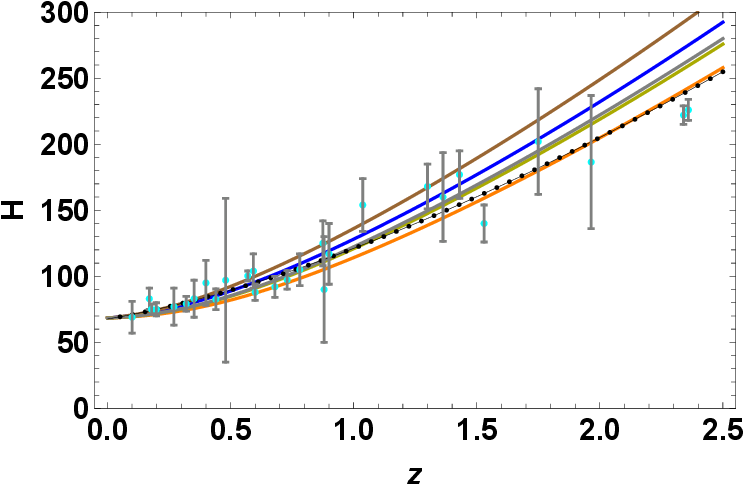}
\caption{Cosmological quantities. The upper-left panel shows the deceleration parameter ({\sf DP}), while the upper-right panel displays the effective ({\sf EoS}) parameter. The lower panels present the Hubble parameter for a phantiverse (left) and a quintaverse (right). In the lower panels, the dotted curve represents the corresponding $\Lambda$CDM prediction. The quantities $x_{3i}$ and $x_{6i}$ denote the initial conditions, while $x_{5i}=-1.500$ is fixed throughout. Note that we have used $H_{0}=68~kms^{-1}Mpc^{-1}$ throught the paper~\cite{aghanim2020}.}\label{figure4}
\end{figure}

\begin{figure}[h!]
\centering
\includegraphics[width=.45\textwidth]{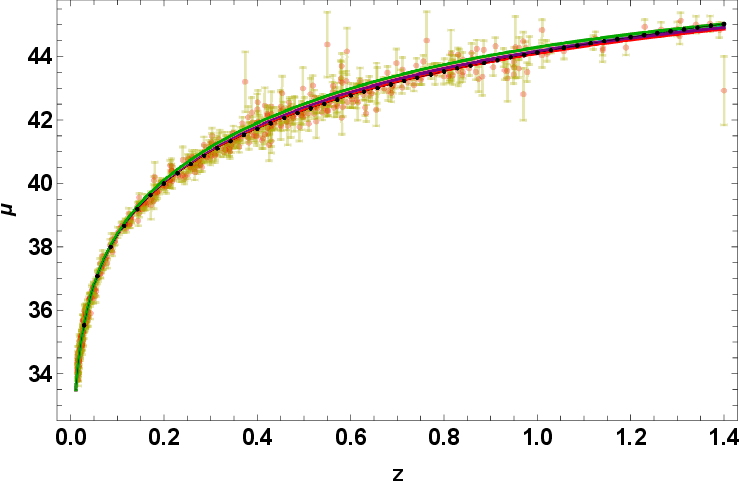}\hspace{2mm}
\includegraphics[width=.45\textwidth]{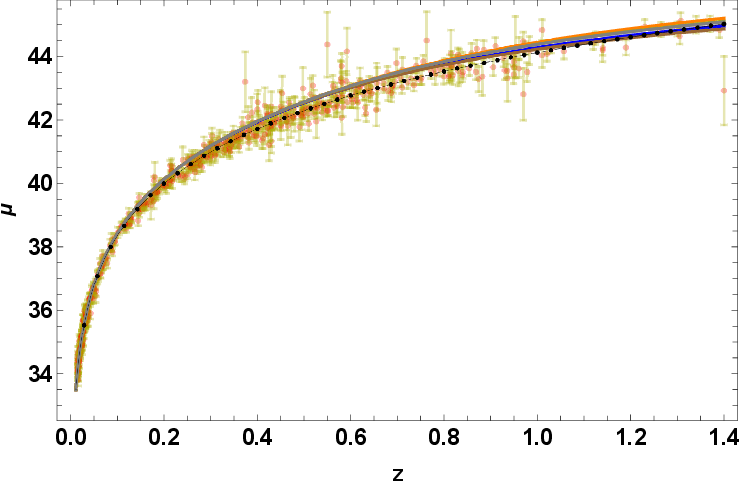}\vspace{2mm}
\includegraphics[width=.45\textwidth]{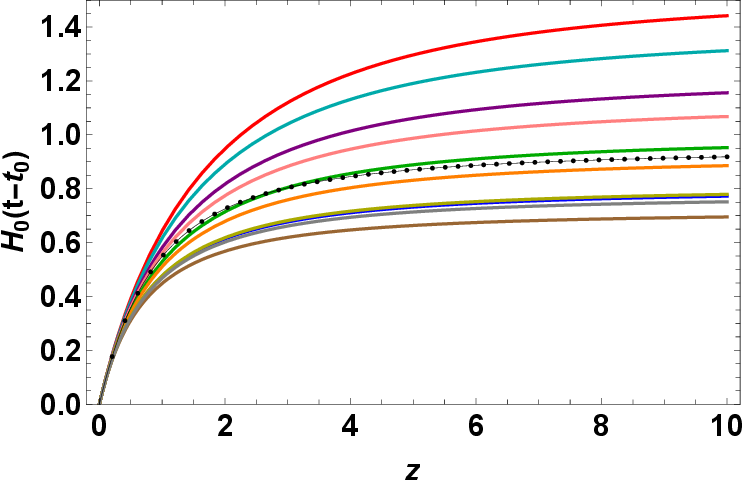}\hspace{2mm}
\includegraphics[width=.45\textwidth]{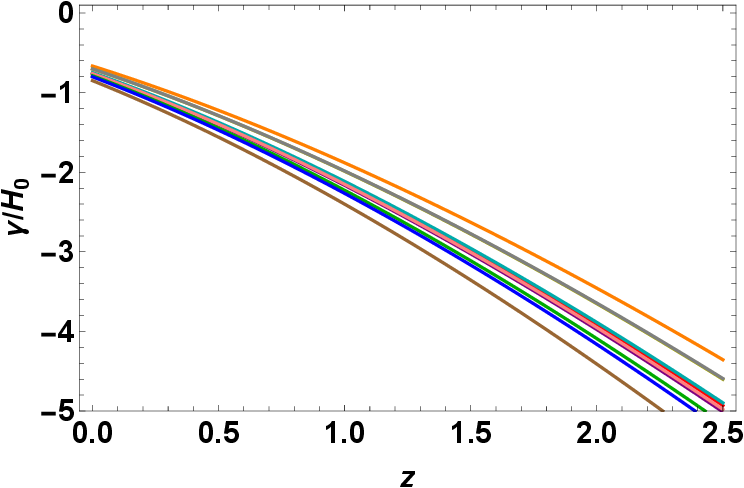}
\caption{The upper panels show the distance modulus for a phantiverse (left) and a quintaverse (right), together with the corresponding observational data. The dotted curve shows the corresponding $\Lambda$CDM prediction. The lower panel illustrates the evolution of the ratio $\gamma/H_0$.}\label{figure5}
\end{figure}

Investigating the behavior of the cosmological quantities is the next step of the present study. The principal quantities of interest are the {\sf DP} $q$, the {\sf EoS} parameter $w_{eff}$, the Hubble parameter $H$ and the distance modulus $\mu$ and the $\gamma$-function, which appears in the affine connection and characterizes the dynamical degree of freedom present in the model. To demonstrate their evolution as functions of the redshift, the system of equations given by eqs.~(\ref{t2.25}),~(\ref{t2.26}) and (\ref{t2.33}) must be reformulated in terms of the redshift variable using the relation $d/dt=-H(z)(1+z)d/dz$. Accordingly, Figs.~\ref{figure4} and~\ref{figure5} are obtained using the following relations

\begin{align}
&q(z)=-1-x_5(z),\label{t2.41.a}\\
&H(z)=H_0\exp\left[-\int_0^z \frac{x_5(z')}{1+z'} \, dz'\right],\label{t2.41.b}\\
&\mu(z)=5\log\left[c(z+1)\int _0^z\frac{dz'}{H(z')}\right]+25,\label{t2.41.c}\\
&\gamma(z)=H(z)x_6(z).\label{t2.41.d}
\end{align}

\noindent Note that Figs.~\ref{figure4} and~\ref{figure5} correspond to the case $x_{5i}=-1.50$. Here, the subscript $i$ denotes the initial values adopted for solving the equations of motion.\\

All figures are presented for $-0.1<\epsilon<+0.1$ in increments of $0.02$. We coin the term ``quintaverse'' to denote a universe with small positive values of $\epsilon$, whereas ``phantiverse'' refers to a universe undergoing accelerated expansion with negative values of $\epsilon$, corresponding to quintessence-like and phantom-like cosmological behaviors, respectively. 

The upper panels of Fig.~\ref{figure4} show that, in a quintaverse, the {\sf DP} and, consequently, $w_{eff}$, exhibit damped oscillatory behavior at late times. In contrast, in a phantiverse, both quantities decrease nearly monotonically. Moreover, in both cases, the initial conditions can be chosen such that the deceleration-to-acceleration transition redshift, as well as the present-day values of this quantity, lie within the observationally accepted ranges~\cite{moresco2016,aghanim2020,myrzakulov2024,Hu2025}. The lower panels of Fig.~\ref{figure4} show the theoretical evolution of the Hubble parameter together with the observational $H(z)$ data and their associated error bars~\cite{farooq2017}, included for comparison. The left and right panels correspond to the phantiverse and quintaverse cases, respectively. The dotted curve represents the corresponding {\sf GR} prediction. As can be seen, the quintaverse is more sensitive to variations in $\epsilon$ than the phantiverse. Moreover, the Hubble parameter in the phantiverse exhibits a closer agreement with the {\sf GR} prediction, indicating smaller deviations from the standard cosmological behavior. Furthermore, in the phantiverse, all curves exhibit a tighter convergence at higher redshifts.\\

The upper panels of Fig.~\ref{figure5} present the theoretical evolution of the distance modulus, separately for the phantiverse (left panel) and the quintaverse (right panel), together with the corresponding observational data~\cite{amanullah2010}. As in the case of the Hubble parameter, the phantiverse exhibits a closer agreement with the corresponding {\sf GR} prediction than the quintaverse. The lower-left panel of Fig.~\ref{figure5} illustrates the dependence of the age of the Universe on the $\epsilon$ parameter. In a phantiverse, the age of the Universe is larger than the prediction of the $\Lambda$CDM model, whereas in a quintaverse, it is smaller. Finally, the lower-right panel of Fig.~\ref{figure5} shows that the $\gamma$-function remains negative and increases for all values of $\epsilon$ as the redshift decreases. The large magnitude of $\gamma$ at higher redshifts may indicate a stronger interaction between the modified geometric sector and the cosmic fluid during the {\sf DM}-dominated era. This interaction weakens toward lower redshifts, which may be associated with the transition toward a more rapid accelerated expansion of the Universe.

\subsection{Further remarks on the matter density variable}\label{numerical-3}

In the present model, the late-time accelerated expansion emerges as an intrinsic geometrical effect of the $f(Q,C)$ sector, without invoking an independent dark energy ({\sf DE}) component. More precisely, unlike {\sf GR}, the present $f(Q,C)$ theory does not contain an Einstein--Hilbert sector. Therefore, the variable $\widetilde {\Omega}_{{\sf m}}$ introduced in this work should be interpreted as a normalized matter variable, constructed to render the cosmological equations autonomous, rather than as the standard cosmological matter density parameter. Within this framework, the condition $\widetilde{\Omega}_{{\sf m}}=0$ indicates the absence, or a negligible contribution, of the matter sector, whereas $\widetilde{\Omega}_{{\sf m}}>0$ together with $w_{{\sf eff}}=0$ identifies a {\sf DM}-dominated dynamical regime. Accordingly, the cosmological evolution, and in particular the onset of accelerated expansion, is characterized by the effective {\sf EoS} parameter, the {\sf DP}, and the stability properties of the cosmological fixed points.

\begin{figure}[h!]
\centering
\includegraphics[width=.446\textwidth]{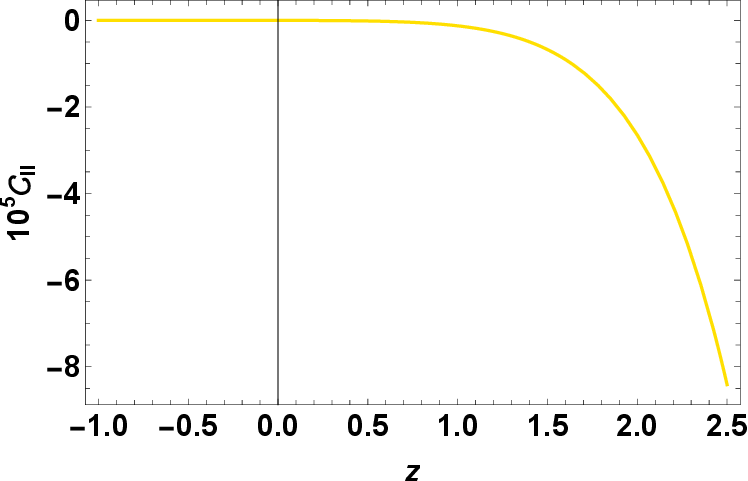}\hspace{2mm}
\includegraphics[width=.45\textwidth]{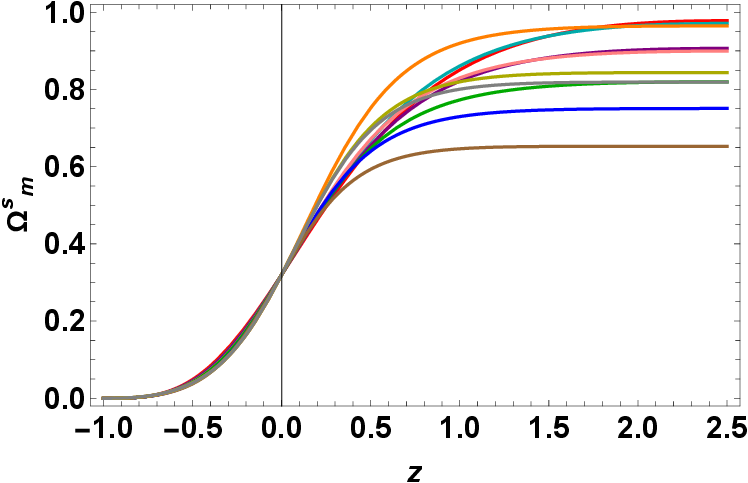}\vspace{2mm}
\caption{Left panel: The bahavior of the $\mathcal{C}_{II}$ parameter corresponding to eq.~(\ref{t2.10}) for $\epsilon=-0.08$, $\eta_1=-0.03$ and $\eta_2=+0.01$. Right panel: The evolution of the standard matter density parameter, defined through the Friedmann equations, for different values of the $\epsilon$ parameter. All curves are plotted such that the present value satisfie $\Omega_{\sf m,0}^s\approx 0.315$~\cite{aghanim2020}.}\label{fig6}
\end{figure}

It is worth mentioning that the contribution of $\mathcal{C}_{II}$ remains effectively negligible throughout the cosmic evolution, since, $x_6=0$ for $P_1$, $P_2$ and $P_3$. Moreover, as discussed previously, $x_6=0$ is also set to zero at the matter point $P_4$.
The evolution of $\mathcal{C}_{II}$ (see eqs.~(\ref{t2.3}) and~(\ref{t2.10})) is shown in the left panel of Fig.~\ref{fig6} for the set of model parameters corresponding to the pink region in Fig.~\ref{fig1}. As can be seen, $\mathcal{C}_{II}$ remains of the order of $10^{-5}$ throughout the entire cosmic evolution, from the {\sf DM}-dominated to the {\sf AE} phase.\\

Therefore, we can safely assume that the {\sf EMT} is conserved, and consequently the matter density evolves as in {\sf GR}, namely $\rho_{\sf m}=\rho_{\sf m,0}(1+z)^3$ for a cosmic fluid with $w=0$, where the subscript ``$0$'' denotes the present value. Hence, the standard matter density parameter, defined as $\Omega_{\sf m}^s=\rho_{\sf m}/\rho_{\sf c}$ where ``c'' denotes the critical density, is obtained as

\begin{align}\label{t2.omega}
\Omega_{\sf m}^s=\Omega_{\sf m,0}^s (1+z)^{3} E_{\sf th}^{-2},
\end{align}

\noindent where $E_{\sf th}=H_{\sf th}/H_0$ is the dimensionless Hubble parameter derived from the background evolution of the underlying theory. In the right panel of Fig.~\ref{fig6}, the evolution of the standard matter density parameter is presented for different values of the $\epsilon$ parameter. All cases exhibit matter domination at early times, with $\Omega_{\sf m}^s$ decreasing toward the present value $\Omega_{\sf m,0}^s\approx 0.315$~\cite{aghanim2020} . This behavior agrees with the standard cosmological picture supported by current observations.

\subsection{Cosmographic and statefinder parameters}\label{numerical-4}

A useful toolkit for characterizing the evolution of the Universe within a given theory of gravity is provided by a set of kinematic quantities defined in terms of successive time derivatives of the scale factor. This approach, commonly referred to as  ``cosmography'', offers a model-independent framework for studying the kinematic evolution of the Universe and for comparing the predictions of different gravity theories with those of the $\Lambda CDM$ model. In the literature, two classes of such diagnostic parameters are widely employed. In the following, we briefly discuss both of them.

\subsubsection{Cosmographyic parameters}\label{numerical-4a}

The cosmographic parameters have their roots in classical mechanics. Their modern formulation in cosmology, together with the terminology of the jerk and snap parameters, was developed by Matt Visser~\cite{visser1,visser2}. Using the definitions of the jerk (j) and snap (s) parameters, one obtains

\begin{align}
&j=\frac{a^{(3)}}{aH^3}=-(z+1) x_5'+x_5 \left(2 x_5+3\right)+1,\label{t2.42}\\
&s=\frac{a^{(4)}}{aH^4}=-3 x_5 (z+1) x_5'-\Big[\big(3 \left(-x_5-1\right)+2\big) \left(x_5+1\right) \left(2 x_5+1\right)\Bigg]\label{t2.43},
\end{align}

\noindent in terms of the phase space variable $x_5$, where the superscript ($i$) on the scale factor denotes its $i$th time derivative, and the prime denotes differentiation with respect to the redshift. For the $\Lambda CDM$ model, one readily obtains
\begin{align}
&j_{\Lambda CDM}=\frac{1}{2} \big[9 (w_{\Lambda } +1) \Omega _{\Lambda } w_{\Lambda }+2,\big],\label{t2.44}\\
&s_{\Lambda CDM}=\frac{1}{4} \bigg\{-9 \Omega _{\Lambda } w_{\Lambda } \Big[w_{\Lambda } \big(3 \Omega _{\Lambda }+3 \left(\Omega _{\Lambda }+2\right) w_{\Lambda }+13\big)+9\Big]-14\bigg\},\label{t2.45}
\end{align}

\noindent where $\Omega _{\Lambda}$ and $w_{\Lambda }=p_{\Lambda }/\rho_{\Lambda }$ denote the density parameter and the {\sf EoS} parameter of the {\sf DE} component, respectively. The jerk parameter remains identically equal to unity throughout the cosmic evolution. In contrast, the snap parameter approaches $-7/2$ during the {\sf DM} era and tends to unity at late times.\\

The left panel of Fig.~\ref{fig7} illustrates the evolutionary trajectories of the Universe in the $j-s$ plane. The dot-dashed curve represents the $\Lambda$CDM evolution during the {\sf DM} era, whereas the dashed curve corresponds to the {\sf DE}-dominated epoch. The characteristic $j_{\Lambda CDM}=1$, is shown as a solid black line. The trajectories originate near the green point with coordinates $(1,-3.5)$ and terminate close to the black point at $(1,1)$. The initial states do not exactly coincide with $(1,-3.5)$, which represents the {\sf DM} fixed point of the $\Lambda CDM$ model. This deviation may indicate that the {\sf DM}-dominated era in the present model has a different nature, possibly arising from the interaction between the matter and geometric sectors, rather than from the standard {\sf DM} solution of $\Lambda CDM$. Considering the colors associated with the selected values of $\epsilon$ (see the upper-left panel of Fig.~\ref{figure4}), we find that the {\sf DM} solution in a quintaverse is closer to the corresponding matter era of the $\Lambda CDM$ model. The jerk parameter remains positive, indicating a decreasing trend of {\sf DP}. The transition from increasing to decreasing values of the jerk parameter reflects the change in the rate of variation of the deceleration parameter. Moreover, the transition of the snap parameter from negative to positive values is consistent with the aforementioned behavior of the jerk parameter.

\subsubsection{Statefinder parameters}\label{numerical-4b}

The statefinder variables introduced in~\cite{sahni}, are calculated as

\begin{align}
&r_{{\sf sf}}=\frac{a^{(3)}}{aH^3}=-(z+1) x_5'+x_5 \left(2 x_5+3\right)+1,\label{t2.46}\\
&s_{{\sf sf}}=\frac{r_{{\sf sf}}-1}{3 \left(q-\frac{1}{2}\right)}=\frac{2}{3} \left[\frac{(z+1) x_5'}{2 x_5+3}-x_5\right]\label{t2.47}.
\end{align}

The subscript ``{\sf sf}'' denotes the statefinder quantities. Therefore, the jerk parameter has the same definition as its cosmographic counterpart (and consequently takes the same value in the $\Lambda CDM$ model). However, the snap parameter is defined differently in the statefinder framework. For the $\Lambda CDM$ model, the corresponding statefinder snap parameter is obtained as\footnote{Note that, the equations in~(\ref{t2.44}),~(\ref{t2.45}) and~(\ref{t2.48}) are valid for cosmological states in which the {\sf EoS} parameter remains constant (see~\cite{sahni}).}

\begin{align}\label{t2.48}
s_{{\sf sf}}^{\Lambda CDM}=1+w_{\Lambda },
\end{align}

\begin{figure}[h!]
\centering
\includegraphics[width=.443\textwidth]{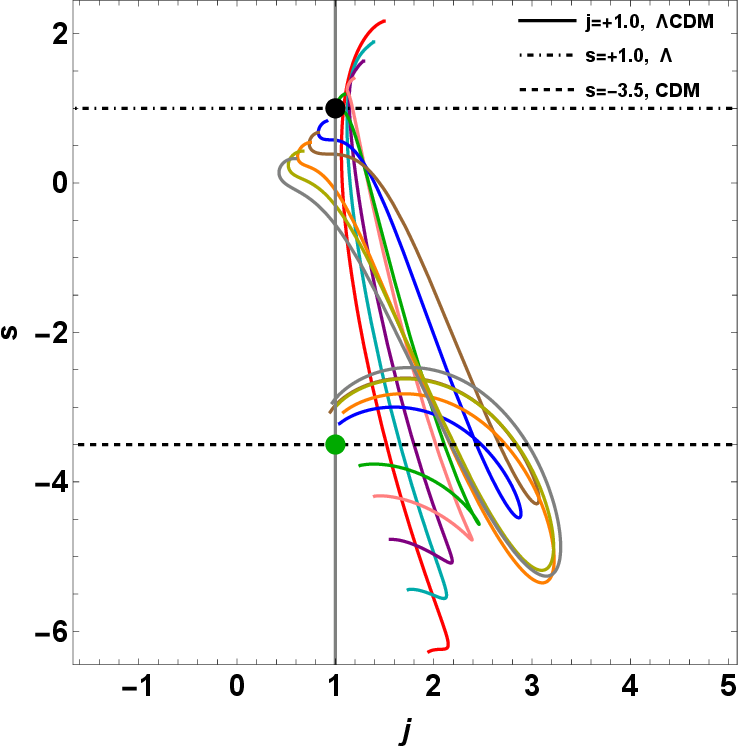}\hspace{2mm}
\includegraphics[width=.455\textwidth]{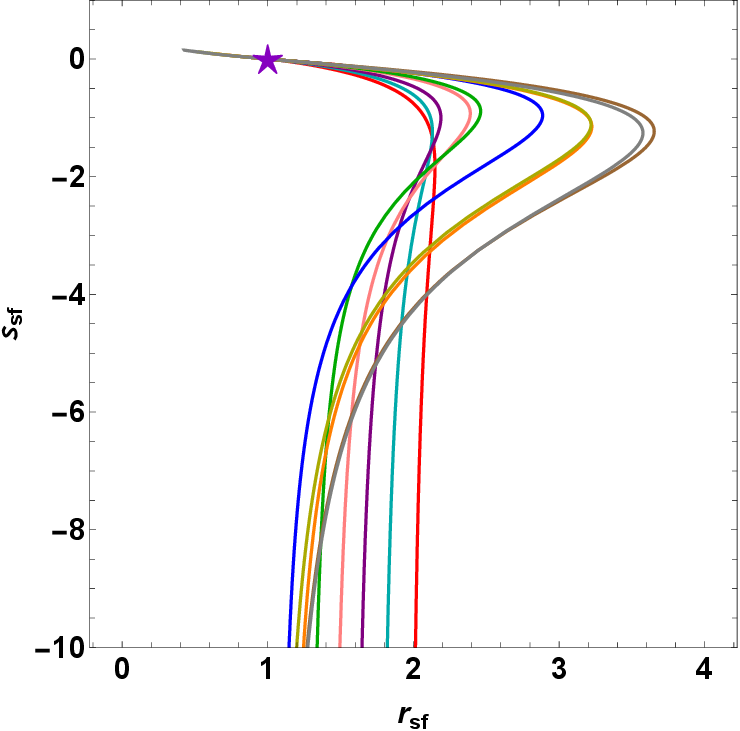}\vspace{2mm}
\caption{The left panel shows the evolutionary trajectories in the cosmographic $j-s$ plane. The green point represents the  {\sf DM}-dominated solution of the $\Lambda CDM$ model, while the black point corresponds to its {\sf DE}-dominated solution. The dot-dashed black line denotes the $s=+1$ line associated with the {\sf DE}-dominated epoch, whereas the dashed line represents the $s=-3.5$ line corresponding to the {\sf DM} era in the $\Lambda CDM$ model. The right panel shows the evolutionary trajectories in the statefinder $r_{{\sf sf}}-s_{{\sf sf}}$ plane. The purple star represents the $\Lambda CDM$ fixed point, corresponding to $(r_{{\sf sf}}, s_{{\sf sf}})=(1,0)$, which remains unchanged throughout the cosmic evolution.}\label{fig7}
\end{figure}

\noindent The right panel of Fig.~\ref{fig7} demonstrates the corresponding trajectories in the statefinder plane. In the statefinder diagnostic approach, the $\Lambda CDM$ model is represented by the fixed point $(r_{{\sf sf}}, s_{{\sf sf}})=(1,0)$. A similar behavior can be observed in this case: the Universe evolves from a state with a jerk parameter close to unity, where an increase-to-decrease transition in its value is again exhibited. Finally, all trajectories approach the $\Lambda CDM$ fixed point (the purple star), indicating that the underlying models become indistinguishable from $\Lambda CDM$ in late times within the statefinder diagnostic.

\section{Connection class $\Gamma_{III}$}\label{type3}

In this section, we investigate the cosmological evolution governed by the third connection class. To keep the discussion concise, only the main results are presented.

\subsection{The {\sf EoMs} and the corresponding equations}\label{III-eom}

\noindent For this connection class, the {\sf EoMs} take the following form:

\begin{align}
&\kappa \rho_{{\sf m}}=\frac{f}{2}+f_Q \left(3 H^2-\frac{Q}{2}\right)-\frac{C f_C}{2}-\frac{3}{2}\frac{\gamma}{a^2}\dot{f}_Q+\left(3 H+\frac{3}{2}\frac{\gamma}{a^2}\right)\dot{f}_C,\label{t3.1}\\
&\kappa p_{{\sf m}}=-\frac{f}{2}+f_Q \left(-3 H^2-2 H'+\frac{Q}{2}\right)+\frac{C f_C}{2}+\frac{1}{2} \left(\frac{\gamma}{a^2} -4 H\right) \dot{f}_Q-\frac{1}{2}\frac{\gamma}{a^2}\dot{f}_C-\ddot{f}_C,\label{t3.2}\\
&\mathcal{C}_{III}=-\frac{3}{2}\frac{\gamma}{a^2}\left[\left(H+2\frac{\dot{\gamma}}{\gamma}\right)\dot{f}_Q+\ddot{f}_Q-\left(H+2\frac{\dot{\gamma}}{\gamma}\right)\dot{f}_C-\ddot{f}_C\right]\label{t3.3},
\end{align}

with

\begin{align}
&Q=-6 H^2+\frac{3}{a^{2}}\big( \gamma  H+ \dot{\gamma}\big),\label{t3.4}\\
&C=6 \left(3 H^2+\dot{H}\right)-\frac{3}{a^{2}}\big( \gamma  H+ \dot{\gamma}\big)\label{t3.5}.\\
\end{align}

\noindent Here, we redefine $x_6$ according to $x_6=\gamma/(Ha^{2})$, while all other variables introduced in~(\ref{t2.7}), and~(\ref{t2.7.1}) remain unchanged\footnote{We also use a $\bar{\Omega}_{\sf m}$ instead of $\widetilde{\Omega}_{\sf m}$ for distinction.}. Consequently, the following autonomous system is obtained 

\begin{align}
&\bar{\Omega}_{\sf m}=\nonumber\\
&\frac{1}{6} \Big[(x_1 \left(12 \eta _1-\eta _3 x_2\right)-\eta _1 x_1^2+\eta _2 \big(6 x_3 \left(x_6+2\right)-x_2^2\big)+3 \eta _3 \left(2 x_2-x_3 x_6\right)+3 x_4 \big(\eta _3 \left(x_6+2\right)-2 \eta _1 x_6\big)\Big],\label{t3.6}\\
&x_{7}=\frac{1}{4 \eta _2}\Bigg\{\bigg[\eta _1\Big(x_1 \left(x_1-8 x_5-12\right)+2 x_4 \left(x_6-4\right)\Big)+\eta_{2}\Big(x_2^2-2 x_3 x_6\Big)\nonumber\\
&+\eta_{3}\Big(x_2 \left(x_1-4 x_5-6\right)+x_3 \left(x_6-4\right)-x_4 x_6\Big)\bigg]-6 w\bigg[\eta_{1}\Big(-\frac{x_1^2}{6}+2 x_1-x_4 x_6\Big)\nonumber\\
&+\eta_{2}\Big(-\frac{x_2^2}{6}+x_3 \left(x_6+2\right)\Big)+\eta_{3}\Big(-\frac{1}{6} \left(x_1-6\right) x_2+x_4+\frac{1}{2} \left(x_4-x_3\right) x_6\Big)\bigg]\Bigg\}-\frac{\eta_3}{2\eta_2}\frac{\ddot{Q}}{H^4},\label{t3.7}\\
&\mathcal{C}_{III}=\frac{3}{2}\Bigg\{\eta_{1} \Big[x_4 \big(-4 \frac{ \dot{\gamma}}{a^2}-2 H^2 x_6\big)H^2-2 x_6 \ddot{Q})\Big]-\eta_{2}\Big[4x_3 \frac{\dot{\gamma}}{a^2}+2  x_6 \left(x_3+x_7\right)H^2\Big]H^{2}\nonumber\\
&+\eta_{3}\Big[2\left(x_4-x_3\right)H^2\frac{ \dot{\gamma}}{a^2}-x_6 \left(x_3-x_4+x_7\right)H^4 +x_6 \ddot{Q}\Big]\Bigg\}H,\label{t3.8}
\end{align}

\noindent for eqs.~(\ref{t3.1}),~(\ref{t3.2}) and~(\ref{t3.3}), respectively. Furthermore, by adopting the same $\gamma$-function as in the previous sections, we obtain

\begin{align}
&\dot{\gamma}=n x_6 a^2 H^2,\label{t3.9}\\
&\ddot{\gamma}=\frac{ a^2}{3} H^3 \big(2 x_1+x_4-3 x_5 \left(x_6-4\right)+12\big)-H \dot{\gamma},\label{t3.10}\\
&\dddot{\gamma}=\frac{ a^2}{3}\bigg\{2 H^4 \Big(x_1 \left(x_5+2\right)+2 \big(x_4+3 x_5 \left(x_5+5\right)+6\big)\Big)-3 H (x_6-4) \ddot{H}+\ddot{Q}\bigg\}-H \big(2 H x_5 \dot{\gamma}+\ddot{\gamma}\big),\label{t3.11}\\
&\ddot{Q}=\frac{1}{2}\bigg\{x_6 \Big[6 n \big((l+4) m+2\big)+6 x_5 \big(n(3 m+2)-4\big)+(n+1) x_3+(n+1) x_4\Big]\nonumber\\
&-4 \Big[x_3+3 x_4+x_1 \left(x_5+2\right)+6 \big(x_5^2+x_5+2\big)\Big]\bigg\}H^4,\label{t3.12}
\end{align}

\noindent together with the following constraint equations

\begin{align}
&x_1+6=3 (n+1) x_6,\label{t3.13}\\
&2 x_1+x_4+12 (x_5+1)=3 x_6 \big[n (m+x_5+2)+x_5\big].\label{t3.14}
\end{align}

\noindent Note that eqs.~(\ref{t2.11}) and eq.~(\ref{t2.12}) remain unchanged. Consequently, the autonomous dynamical system takes the following form

\begin{align}
&\frac{dx_{3}}{dN}=\frac{1}{4 \eta_2}\left(\eta_1\mathcal{A}_1+\eta_2\mathcal{A}_2+\eta_3\mathcal{A}_3\right),\label{t3.15}\\
&\mathcal{A}_1=\big(3 (n+1) x_6-6\big)^2-4 \left(2 x_5+3\right) \big(3 (n+1) x_6-6\big)\nonumber\\
&+2 (x_6-4) \Big[3 \big(n \left(m+x_5+2\right)+x_5\big)x_6 -2 \big(3 (n+1) x_6-6\big)-12 \left(x_5+1\right)\Big],\nonumber\\
&\mathcal{A}_2=\big(6 \left(x_5+3\right)-3 (n+1) x_6\big)^2-2 x_3 \left(6 x_5+x_6\right),\nonumber\\
&\mathcal{A}_3=-216 +\bigg\{6 \big[n (-l m+2 m+17)+7\big]-n x_3-3  \Big[n \big(m (n+2)+3 n+4\big)-1\Big]x_6\bigg\}x_6 \nonumber\\
&+3 x_5 \Big[-96+\big(-6 m n-(n+1) (n+2) x_6+26 n+42\big)x_6\Big], \nonumber\\
&\frac{dx_{5}}{dN}=\frac{1}{6} \Big[x_3-12 x_5 \left(x_5+3\right)+3 \big(n \left(m+x_5\right)+x_5-2\big)x_6\Big],\label{t3.16}\\
&\frac{dx_{6}}{dN}=x_6 \left(n-x_5-2\right).\label{t3.17}
\end{align}

\noindent Again, the expression for$dx_{6}/dN$ reveals two branches of solutions, namely $x_6=0$ and $x_5=n-2$. To ensure the existance of a {\sf DM} solution we set $n=1/2$ which implies $m=-1/2$ and $l=-3/2$. Five critical-point solutions are then obtained, two of which are presented in Table~\ref{tab2}. The remaining three have exactly the same properties as $P_1$, $P_2$ and $P_3$ listed in Table~\ref{tab1}. For consistency with the notation $Q_4$ and $P_5$ used in Table~\ref{tab2}, we denote these points by $Q_1$, $Q_2$ and $Q_3$, respectively. Furthermore, the matter points in this case exhibit different properties from their counterparts obtained for the class-$II$ connection.\\\\\\\\\\\\

\begin{table}[h!]
\centering
\caption{Fixed-point solutions for the system described by eqs.~(\ref{t3.15})-(\ref{t3.17}) for $n=1/2$, $m=-1/2$, and $l=-3/2$.}
\begin{adjustbox}{width=1\textwidth}
\begin{tabular}{l @{\hskip 0.1in}l @{\hskip 0.1in} l @{\hskip 0.1in}l @{\hskip 0.1in}@{\hskip 0.1in}ll}\hline\hline

Fixed point     &$(x_3,x_5,x_6)$  &$\bar{\Omega}_{\sf m}$      &$w_{eff}$\\[0.5 ex]
\hline
$Q_4$    & $\left(9\frac{\mathcal{\bar {B}}_2+7 \eta _1+21 \eta _2-14 \eta _3}{\eta _1+\eta _2-\eta _3},-\frac{3}{2},\frac{2}{3}\frac{\mathcal{\bar {B}}_2+10 \eta _1+24 \eta _2-17 \eta _3}{\eta _1+\eta _2- \eta _3}\right)$              
                        & $\bar{\Omega}_4$
            & $0$ \\[0.75 ex]
$Q_5$    & $\left(9\frac{-\mathcal{\bar {B}}_2+7 \eta _1+21 \eta _2-14 \eta _3}{\eta _1+\eta _2-\eta _3},-\frac{3}{2},\frac{1}{3}\frac{-\mathcal{\bar {B}}_2+10 \eta _1+24 \eta _2-17\eta _3}{\eta _1+\eta _2-\eta _3}\right)$              
                        & $\bar{\Omega}_5$& $0$  \\[0.75 ex]
\hline\hline
\end{tabular}
\end{adjustbox}
\label{tab2}
\end{table}

\noindent Where in Table~\ref{tab2} we used 

\begin{align}
&\mathcal{\bar {B}}_1=\sqrt{\eta _2 \left(\eta _1+4 \eta _2-2 \eta _3\right)},\label{t3.18}\\
&\mathcal{\bar {B}}_2=\sqrt{64 \eta _1^2+309 \eta _2 \eta _1-232 \eta _3 \eta _1+441 \eta _2^2+217 \eta _3^2-609 \eta _2 \eta _3},\label{t3.19}\\
&\bar{\Omega}_4=\frac{45}{2 \left(\eta _1+\eta _2- \eta _3\right)}\Big[2 \eta _1 \left(2 \mathcal{B}_2+79 \eta _2-57 \eta _3\right)+10 \eta _2 \left(\mathcal{B}_2+21 \eta _2\right)+32 \eta _1^2+103 \eta _3^2-\eta _3 \left(7 \mathcal{B}_2+292 \eta _2\right)\Big],\label{t3.20}\\
&\bar{\Omega}_5=\frac{45}{2 \left(\eta _1+\eta _2- \eta _3\right)}\Big[-2 \eta _1 \left(2 \mathcal{B}_2-79 \eta _2+57 \eta _3\right)+32 \eta _1^2+210 \eta _2^2+\eta _3 \left(7 \mathcal{B}_2+103 \eta _3\right)-2 \eta _2 \left(5 \mathcal{B}_2+146 \eta _3\right)\Big].\label{t3.21}
\end{align}

\subsection{The $\epsilon$-parameterization }\label{III-epsilon}

Similar to the procedure followed in Subsec.~\ref{subsec2.6}, we apply the $\epsilon$-parameterization to the present theory. We again discard the point $Q_1$(=$P_1$), while identifying $Q_2/Q_3$ as the solutions corresponding to an ${\sf AE}$ era. Here, $Q_2$ and $Q_3$ are parameterized for opposite signs of $\eta_2$, and the same conditions given in~(\ref{t2.31.a}) and~(\ref{t2.31.b}) apply when identifying $Q_2=P_2$ and $Q_3=P_3$, respectively. Consequently, we obtain results similar to those discussed in Subsec.~\ref{numerical}. In particular, $Q_1$, $Q_2$ and $Q_3$ exist in both theories, corresponding to the use of either the class-$II$ or class-$III$ connection, and exhibit the same properties. This common behavior indicates the robustness and universality of these solutions.

\subsection{The conservation of the {\sf EMT} }\label{III-conserv}

An important difference arises between the theories employing the connection class $\Gamma_{III}$ and those based on the class $\Gamma_{II}$. Substituting eq.~(\ref{t3.9}) into eq.~(\ref{t3.8}) shows that $x_6=0$ leads to $\mathcal{C}_{III}=0$. Therefore, the conservation of the {\sf EMT} is satisfied at the points $Q_1$, $Q_2$ and $Q_3$. If $Q_4$ is identified as the solution corresponding to the {\sf DM} phase, $\mathcal{C}_{III}$ vanishes for the following parameter choices

\begin{align}\label{t3.22}
\mathcal{C}_{III}^{Q_4}=0~~~\to~~~~~~~~~\eta_2=0,~~~\eta_2=\frac{4 \eta_1}{9 (\epsilon -2)^2},~~~\eta_3=\frac{4 \eta_1}{(3 \epsilon +2)^2}.
\end{align}

\noindent However, for all the above cases, one obtains $\bar{\Omega}_4=0$. The same conclusion is reached when $Q_5$ is identified as the {\sf DM} point. This indicates that, in class $\Gamma_{III}$, the conservation of the of the {\sf EMT} cannot be satisfied simultaneously with the condition $\bar{\Omega}_{\sf m}>0$. Indeed, one obtains 

\begin{align}\label{t3.23}
\mathcal{C}_{III}^{Q_4}=-9H^5,~~~~~\to~~~~\dot{\rho}_{\sf m}+3H\rho_{\sf m}-9H^5=0
\end{align}

\noindent at $Q_4$ for $\bar{\Omega}_4=1$. On the other hand, at the {\sf DM} fixed point, $x_5=-3/2$ which yields $H=H_0 (z+1)^{3/2}$ upon imposing the integration condition $H(0)=H_0$. Substituting this result into the right-hand side of eq.~(\ref{t3.23}) and integrating with respect to redshift, we obtain the following expression for the matter density

\begin{align}\label{t3.24}
\rho_{\sf m}=\left(\rho_{\sf m,0}-3 H_0^4\right)(z+1)^3 +3H_0^4 (z+1)^6,
\end{align}

\noindent in terms of the redshift. We see that the solution~(\ref{t3.24}) consists of contributions associated with both stiff and pressureless matter. Hence, within the class $\Gamma_{III}$, the {\sf DM} phase emerges as an effective cosmological regime resulting from the interplay between the geometrical sector and the matter contributions, including both pressureless and stiff components. This demonstrates the intrinsically coupled nature of geometry and matter in the realization of the effective {\sf DM} phase.

\subsection{Phase-space portraits}\label{III-portraits}

Rewriting eqs.~(\ref{t3.15})-(\ref{t3.17}) in terms of the Poincar\'e coordinates, we obtain

\begin{align}\label{t3.25}
\frac{d X_{i}}{dN}=g_i(X_3,X_5,X_6),~~~~~~i=3,5,6,
\end{align}

\noindent where the functions $g_i(X_3,X_5,X_6)$ are specified in the Appendix by eqs.~(\ref{app5})-(\ref{app7}). Fig.~\ref{fig8} presents two representative examples. The upper panels show a quintaverse for $\eta_1=-0.92$, $\eta_2=-0.04$ and $\epsilon=0.04$, displayed from two different viewpoints. For the lower panel, we use $\eta_1=-0.93$, $\eta_2=-0.03$ and $\epsilon=-0.04$, thereby obtaining a phase-space portrait corresponding to a phantiverse. As before, the green and black points indicate the points $Q_4$ and $Q_2$, corresponding to the {\sf DM} and {\sf AE} solutions, respectively. The gray and red points represent $Q_1$ and $Q_3$, respectively, while the star-shaped markers indicate the fixed-point solutions at infinity. One sees that the main features of the phase-space structure remain essentially unchanged compared with those observed in the upper panels of Fig.~\ref{fig3}.

\begin{figure}[h!]
\centering
\includegraphics[width=.45\textwidth]{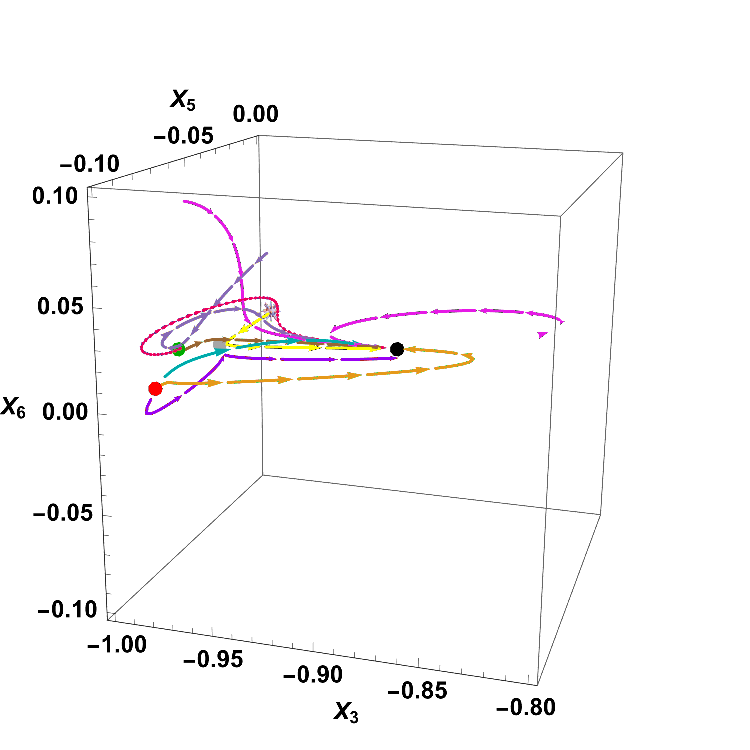}\hspace{2mm}
\includegraphics[width=.45\textwidth]{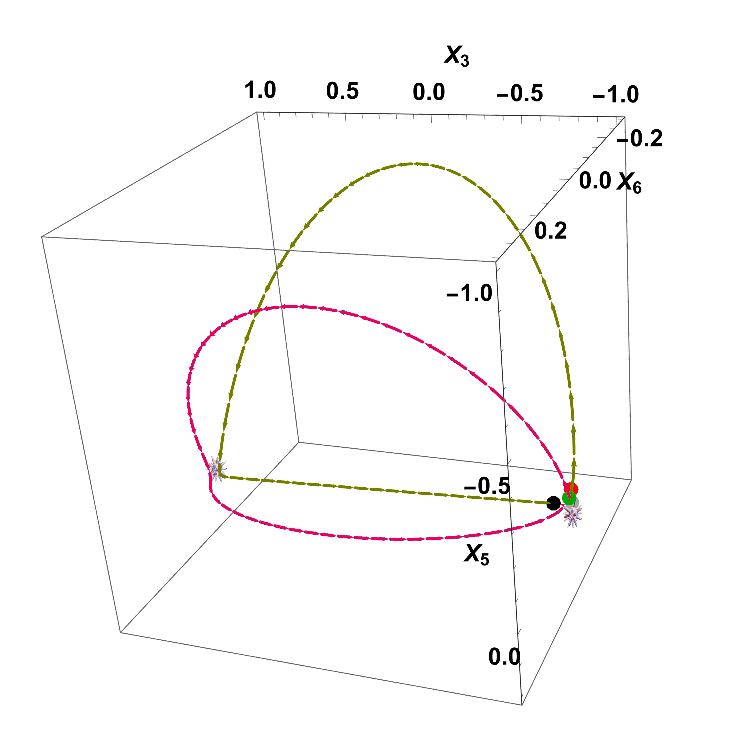}\vspace{2mm}
\includegraphics[width=.45\textwidth]{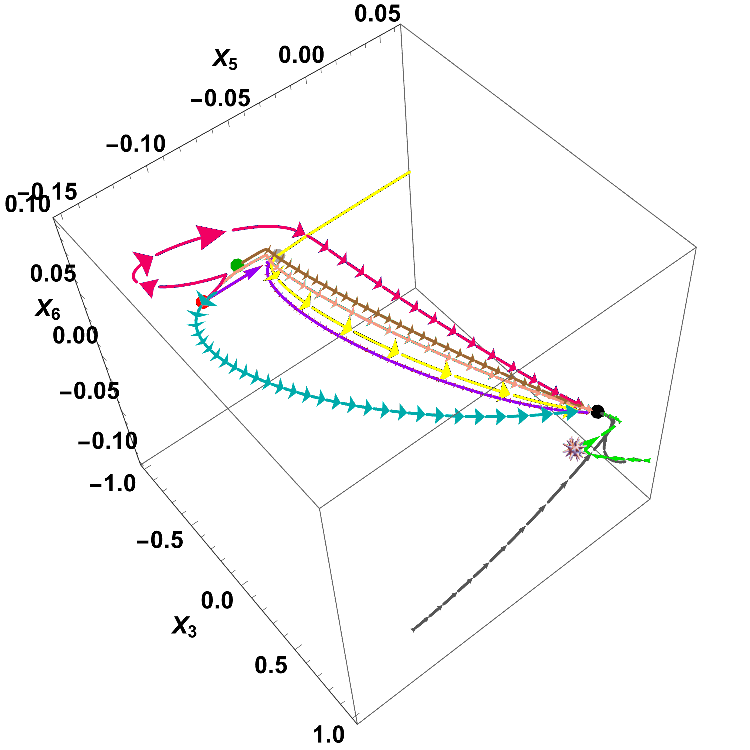}
\caption{Phase-space portraits in the Poincar\'e coordinates for the class $III$ connection. The upper panels show the trajectories for a quintaverse with $\epsilon=+0.04$. The upper-left panel provides a focused view, while the upper-right panel displays selected trajectories that pass through one of the fixed-point solutions at infinity before reaching the black point corresponding to the {\sf AE} solution. The lower panel shows the corresponding phase-space portrait for a phantiverse with $\epsilon=-0.04$.}\label{fig8}
\end{figure}

\subsection{Cosmological quantities}\label{III-quantities}

\begin{figure}[h!]
\centering
\includegraphics[width=.45\textwidth]{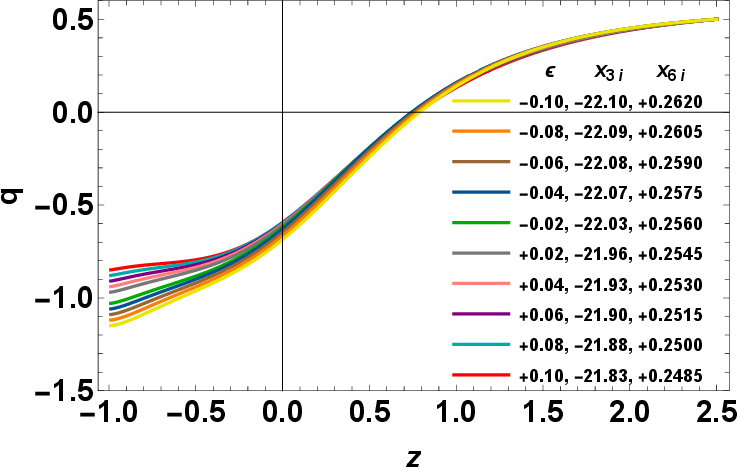}\hspace{2mm}
\includegraphics[width=.455\textwidth]{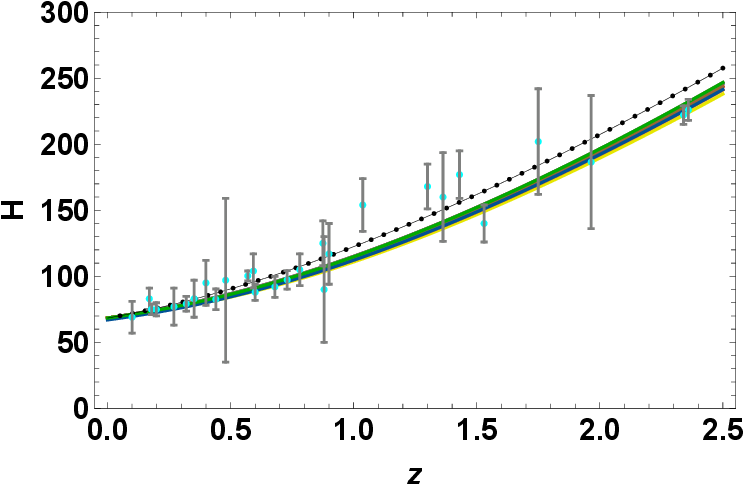}\vspace{2mm}
\includegraphics[width=.44\textwidth]{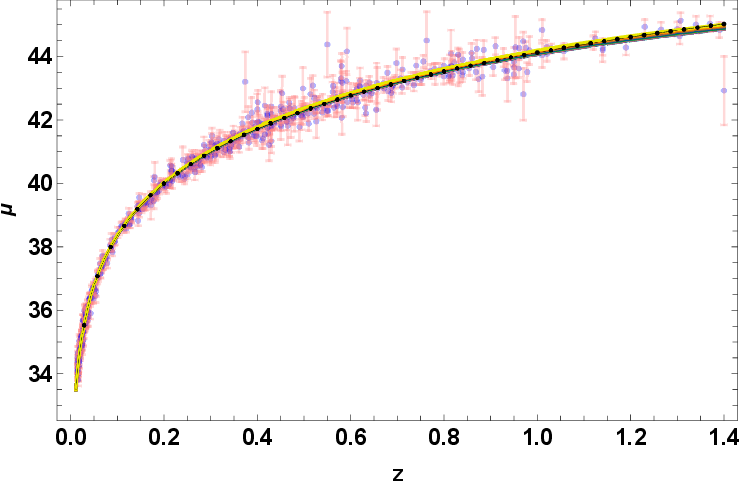}\hspace{2mm}
\includegraphics[width=.45\textwidth]{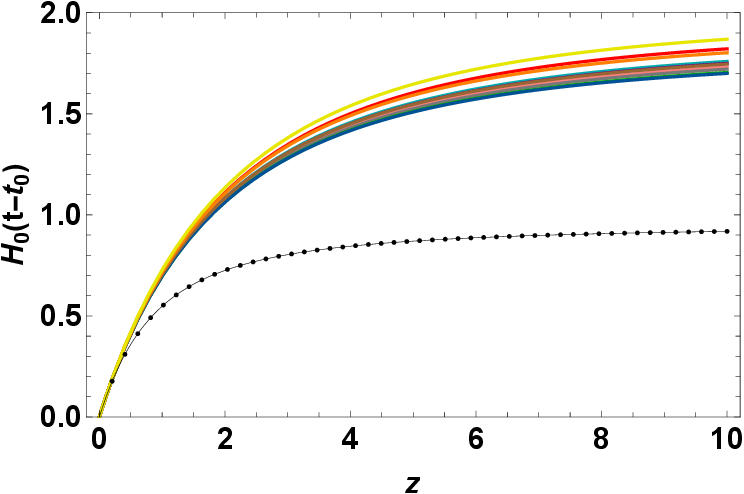}\vspace{2mm}
\includegraphics[width=.46\textwidth]{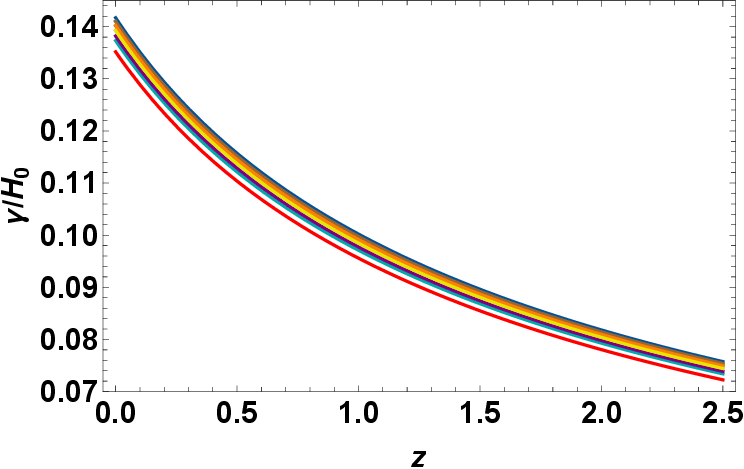}
\caption{Behavior of the cosmological quantities for the class $III$ connection.}\label{fig9}
\end{figure}

Fig.~\ref{fig9} illustrates the background cosmological quantities obtained using the relations~(\ref{t2.41.a})-(\ref{t2.41.d}).  In this case, the initial values have been chosen so as to achieve the best consistency between the distance-modulus curve and its corresponding $\Lambda$CDM counterpart. As can be seen, with these initial values, the Hubble parameter curve passes through a larger number of observational data points than the corresponding $\Lambda$CDM curve.\footnote{A dedicated parameter-constraining analysis can, of course, be performed to determine the best-fit values of the model parameters.} The middle-right panel demonstrates that the underlying theory predicts a larger age of the Universe for all values of $\epsilon$. This may be attributed to the unusual nature of the matter content in this cosmological scenario, in which a stiff fluid dominates over the other components at high redshifts (see solution~(\ref{t3.24})). The lower panel shows that the $\gamma$-function remains positive throughout the evolution and increases from the {\sf DM} era toward the {\sf AE} era.

\subsection{Cosmographic and statefinder parameters}\label{III-cosmographic}

\begin{figure}[h!]
\centering
\includegraphics[width=.443\textwidth]{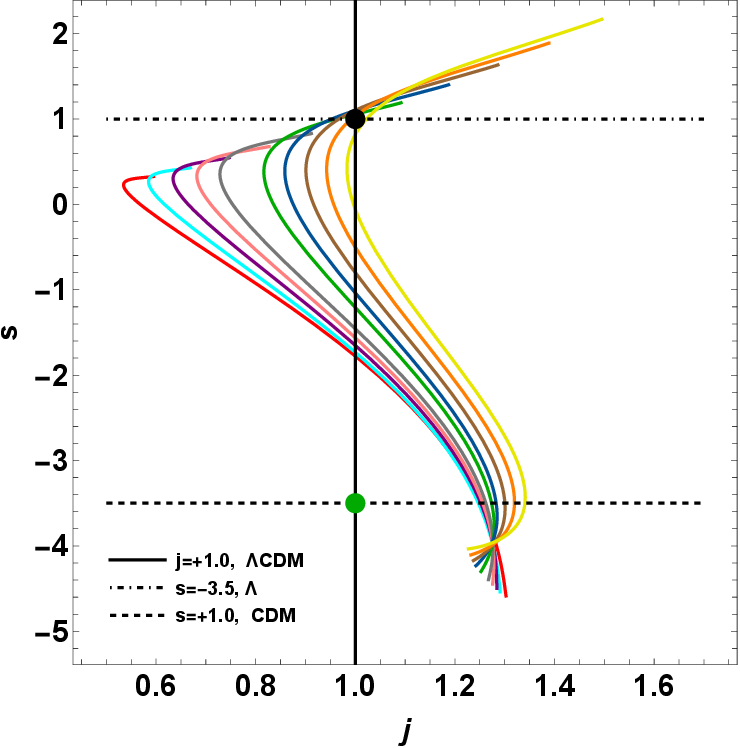}\hspace{2mm}
\includegraphics[width=.47\textwidth]{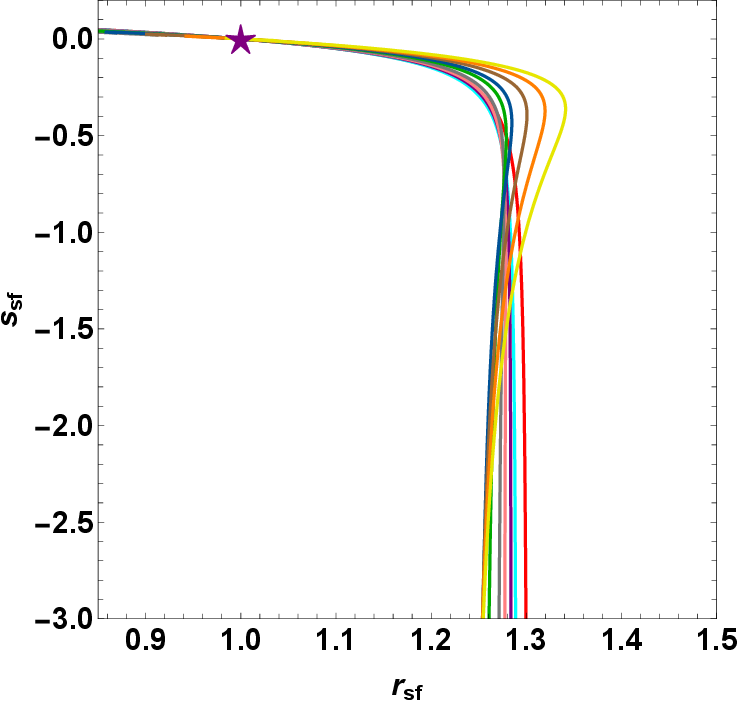}\vspace{2mm}
\caption{Evolution of the cosmographic and statefinder quantities for the class-$III$ connection. The left panel shows the cosmographic parameters, while the right panel displays the statefinder variables.}\label{fig10}
\end{figure}

Figure~\ref{fig10} presents the corresponding trajectories in the cosmographic parameter space (left panel) and the statefinder space (right panel). From the left panel, it is evident that the {\sf DM} phase differs substantially from its $\Lambda$CDM counterpart, as all trajectories originate far from the {\sf DM} fixed point of the $\Lambda$CDM model (green point). This behavior is also consistent with the non-conservation of the {\sf EMT} arising in the class  $\Gamma_{III}$ (see Subsec.~\ref{III-conserv}). 

\section{Concluding remarks}\label{conclusion}

n this work, we explored the evolution of the Uiniverse in $f(Q,C)=\eta_{1} Q^2+\eta_{2} C^2+\eta_{3} CQ$ gravity within a spatially flat FLRW spacetime. As is well known, this theory admits two corresponding non-trivial classes of affine connections\footnote{Apart from the trivial case determined by the relations~(\ref{casesi}), which yields the {\sf $f(\mathbb T)$} field equations and is referred to as the ''coincident'' gauge.}. Throughout this paper, we denote these two classes by $\Gamma_{II}$ and $\Gamma_{III}$, respectively. Hereafter, we refer to the theories employing $\Gamma_{II}$ and $\Gamma_{III}$ as the ``former theory'' and ``the latter theory,'' respectively. Both classes of connections contain a degree of freedom, represented by the $\gamma$-function in the relations~(\ref{casesii}) and~(\ref{casesiii}). We adopt a power-law dependence of the $\gamma$-function on the scale factor. To provide a unified overview of the analysis presented in this work, we first discuss the results that are shared by both connection classes and then summarize the findings for each theory separately in the following sections.\\

\noindent {\bf The main picture}\\

Our analysis of the cosmological dynamics of $f(Q,C)$ gravity was based on a set of dimensionless variables through which the equations of motion were recast as a first-order dynamical system. As discussed in Subsects.~\ref{subsec2.4} and~\ref{III-eom}, this dynamical system is characterized by three independent variables, two of which are associated with the effective {\sf EoS} parameter and the $\gamma$-function. \\

These systems of equations admit five finite fixed-point solutions. We identified a universally stable fixed point\footnote{Here, ``universal'' refers to the fact that the stability of the {\sf AE} solution is maintained for all $\epsilon<2/3$, encompassing the observationally favored values of the effective {\sf EoS} parameter of the Universe.} corresponding to an {\sf AE} era with $w_{\sf eff}=-1+\epsilon$. Depending on the sign of $\eta_2$, this solution is represented by either $P_2$ or $P_3$ in the former theory and by either $Q_2$ or $Q_3$ in the latter theory. We also identified a saddle fixed point corresponding to an effective {\sf DM} era, characterized by $w_{\sf eff}=0$ and $\widetilde{\Omega}_{\sf m}\approx1$ in the former theory ($\bar{\Omega}_{\sf m}\approx1$ in the latter theory), represented by $P_4$ and $Q_4$, respectively. Here, $\epsilon$ characterizes the deviation of the effective {\sf EoS} parameter from $-1$ and is related to $\eta_3$ through $\eta _3=\frac{1}{8} (4 \eta _1-3 \eta _2 (\epsilon -2) (3 \epsilon +2))$.

We referred to this prescription as the $\epsilon$-parameterization and denoted the corresponding cosmological scenarios with $\epsilon<0$ and $\epsilon>0$ as the ``phantiverse'' and ``quintaverse,'' respectively. The remaining finite fixed points include a solution with $w_{\sf eff}=-1/2$, a solution with $w_{\sf eff}=1/3-\epsilon$, corresponding to an effective ``pseudo-ultra-relativistic'' fluid with $\widetilde{\Omega}_{\sf m}=0$ ($\bar{\Omega}_{\sf m}=0$), and another solution with $w_{\sf eff}=0$ and a broad range of $\widetilde{\Omega}_{\sf m}$ values in the former theory ($\bar{\Omega}_{\sf m}$ in the latter theory). In the latter case, $0<\widetilde{\Omega}_{\sf m}<1$ or $0<\bar{\Omega}_{\sf m}<1$ can occur for certain parameter values; such cases were excluded from the present analysis. We further found that the conservation of the {\sf EMT} is automatically satisfied at all fixed points except the matter points. Requiring {\sf EMT} conservation at the matter points, together with the condition $\widetilde{\Omega}_{\sf m}\approx1$ ($\bar{\Omega}_{\sf m}\approx1$), may allow us to constrain the free parameters $\eta_i$ such that both conditions are simultaneously satisfied at the corresponding matter point. Finally, our analysis of the critical points at infinity revealed two additional saddle fixed points for which $w_{\sf eff}$ asymptotically approaches $-1$.\\

Furthermore, we analyzed the phase-space diagrams in the Poincar'e coordinates and examined the evolution of various cosmological quantities, including the {\sf DP}, the Hubble parameter, the distance modulus, and the predicted age of the Universe. We also investigated the evolution of the $\gamma$-function for the range $-0.1<\epsilon<+0.1$ with an increment of $\Delta\epsilon=0.02$. Finally, we examined the former and latter theories in terms of the cosmographic and statefinder variables. Based on these analyses, the following points should also be emphasized.

\begin{itemize}
\item
One might be tempted to consider cases in which one or two of the $\eta_i$ vanish, thereby isolating the effects of the individual $C^2$- or $Q^2$-terms. However, our numerical results, presented in the tables of Figs.~\ref{fig1} and~\ref{fig2}, demonstrate that none of the $\eta_i$ can be consistently set to zero.

\item 
Substituting $C=R-Q$ (see Eq.~(\ref{t2.6})) into the function $f(Q,C)$ given by Eq.~(\ref{Lagrangian}) allows us to express the Lagrangian in terms of the Ricci scalar as $f(R,Q)=\eta _2 R^2+\left(\eta _1+\eta _2-\eta _3\right) Q^2+\left(\eta _3-2 \eta _2\right) Q R$. This representation shows that the original $f(Q,C)$ model can be interpreted as a correction to pure $R^2$ gravity. In particular, the $R^2$ contribution originates solely from the boundary term $C^2$. We also emphasize that the resulting $QR$ term is fundamentally different from the Einstein-Hilbert term.

\item In modified-gravity theories, the formulation of the corresponding dynamical system often involves a non-standard matter variable whose definition, and in some cases dimensionality, differs from that of the standard matter density parameter, which we denote by $\Omega^s=8\pi G\rho/(3H^2)$. For example, in $f(R)$ gravity, one commonly defines~\cite{amendola2007} $\Omega^{f(R)}=8\pi G\rho/3H^2F(R)$, where $F(R)=df(R)/dR$. In our analysis, we referred to the quantity defined in Eq.~(\ref{t2.7.1}) as the ``normalized matter variable'' to emphasize that it should not be identified with the physical matter density parameter. Consequently, it cannot be directly employed as an observational matter density parameter, nor can its magnitude alone be interpreted as a direct measure of matter domination. Nevertheless, its vanishing or positivity still indicates, respectively, the absence or presence of the matter contribution. We therefore used this property in identifying the matter-dominated solutions and imposed the condition that the normalized matter variable equals unity at the fixed points $P_4$ and $Q_4$.
\end{itemize}

\noindent {\bf Results for the former theory}\\

Our analysis of the former theory led to the following specific results.

\begin{itemize}

\item
Based on Eq.~(\ref{t2.26}), one branch of solutions is characterized by $x_5=n$. From the definitions in Eq.~(\ref{t2.7}), we found that choosing $n=-3/2$ ensures the existence of an effective {\sf DM} era. Accordingly, we set $n\left(=\frac{ad\gamma}{\gamma da}\right)=-3/2$ which yields $\gamma=\gamma_0a^{-3/2}$ for the former theory.

\item
The approach in which the two conditions, $\widetilde{\Omega}_{\sf m}\approx1$ and $\mathcal{C}_{II}=0$, are simultaneously satisfied works consistently for several pairs of $(\eta_1,\eta_2)$. As shown in the left panel of Fig.~\ref{fig6}, $\mathcal{C}_{II}\approx10^{-5}$ for the values of $\eta_i$ associated with the pink region in the table of Fig.~\ref{fig6}. Consequently, different sets of fixed-point solutions are obtained for different pairs of $(\eta_1,\eta_2)$, as summarized in the tables of Figs.~\ref{fig1} and~\ref{fig2}.

\item
Based on the behavior of the phase-space trajectories, we found that the sequences $P_4\to P_2$ and $P_3\to P_4\to P_2$ are dynamically allowed. The former sequence may indicate that the effective {\sf DM} era corresponds to an asymptotic solution, whereas in the latter sequence it occurs as a finite-time solution. As it earlier pointed out, the point $P_3$ corresponds to the pseudo-ultra-relativistic fluid.

\item
The right panel of Fig.~\ref{fig6} shows that the standard matter density parameter behaves consistently with both physical expectations and the conceptual picture of cosmic evolution. It decreases from values close to unity toward values consistent with the present observationally favored matter density over the range $-0.1<\epsilon<+0.1$. This behavior is consistent with imposing the condition $\widetilde{\Omega}_{\sf m}=1$ at the effective {\sf DM} fixed point $P_4$.

\item
Upon considering the aforementioned cosmological quantities, we found that the initial conditions and the model parameters $\eta_i$ can be chosen such that the predicted transition redshift from decelerated to accelerated expansion, $z_{\rm trans}\simeq0.6$~\cite{rania2015,muthukrishna2016,aghanim2020}, as well as the present value of the {\sf DP}, $q_0\simeq-0.53$, are consistent with the observational constraints~\cite{aghanim2020,ray2026}\footnote{We do not consider the recent studies reporting values of $q_0$ that are inconsistent with the results of the ``Planck 2018 results''~\cite{verdugo2026,sah2026}.}. We found that the Hubble parameter and the distance modulus in the phantiverse exhibit smaller deviations from their $\Lambda$CDM counterparts. Nevertheless, a likelihood analysis is required to determine the best-fit values of the model parameters. We also found that the age of the Universe in the phantiverse is larger than the corresponding $\Lambda$CDM prediction, whereas that in the quintaverse is smaller. Furthermore, the $\gamma$-function remains negative throughout the cosmic evolution in this theory, while its magnitude increases with redshift. In particular, the existence of the effective {\sf DM} era is associated with large negative values of the $\gamma$-function.

\item
Investigating the cosmographic and statefinder variables, we found that the jerk parameter, which has the same definition in both formalisms, remains positive throughout the cosmic evolution and exhibits an increasing-then-decreasing behavior. The evolution of the cosmographic variables shows that the trajectories originate in the vicinity of the $\Lambda$CDM matter point, characterized by $(j,s)=(1,-3.5)$, and subsequently approach the $\Lambda$CDM dark energy point, $(j,s)=(1,1)$. The left panel of Fig.~\ref{fig7} further indicates that the effective {\sf DM} era in the former theory is more consistent with its $\Lambda$CDM counterpart in the quintaverse. From the evolution of the statefinder variables (see the right panel of Fig.~\ref{fig7}), we found that the former theory becomes indistinguishable from the dark energy solution of the $\Lambda$CDM model only at late times.

\end{itemize}

\noindent {\bf Results for the latter theory}\\

For the latter theory, our analysis yielded the following specific results.

\begin{itemize}

\item
In this case, Eq.~(\ref{t3.17}) shows that one class of solutions is characterized by $x_5=n-2$. We therefore found that the existence of an effective {\sf DM} era requires $n=1/2$, yielding $\gamma=\gamma_0a^{1/2}$ for the latter theory.

\item
At the fixed point $Q_4$,, we found that the two conditions $\bar{\Omega}_{\sf m}\approx1$ and $\mathcal{C}_{III}=0$ cannot be simultaneously satisfied at the effective {\sf DM} fixed point $Q_4$. In fact, imposing $\mathcal{C}_{III}=0$ leads to negative values of $\bar{\Omega}_{\sf m}$. We consequently imposed only the latter condition to constrain the values of $\eta_1$ and $\eta_2$. The resulting non-conservation of the {\sf EMT} gives $\mathcal{C}_{III}^{Q_4}=-9H^5$, which, subject to the present condition $\rho_{\sf m}(t_0)=\rho_{{\sf m},0}$, yields $\rho_{\sf m}=\left(\rho_{\sf m,0}-3 H_0^4\right)(z+1)^3 +3H_0^4 (z+1)^6$. This result indicates that, in the latter theory, the cosmic matter that effectively mimics the {\sf DM} component consists of both a pressureless matter contribution and a stiff-fluid contribution. In this sense, the effective {\sf DM} solution in the latter theory provides a manifestation of the matter-geometry coupling.

\item
We again found the sequences $Q_4\to Q_2$ and $Q_3\to Q_4\to Q_2$ in the phase-space trajectories. The point $Q_3$ plays the same role as $P_3$. Labeling the fixed points at infinity by $S_{\pm}$, located at $(\pm1,0,0)$, respectively, we also identified the sequence $Q_4\to S_{+}\to Q_2$. This sequence can be understood from the fact that $Q_4$ lies on the same side of the phase space as $S_{-}$ (see the upper-right panel of Fig.~\ref{fig8}).

\item
The standard matter density parameter does not remain normalized in this theory because the {\sf EMT} is not conserved.

\item
By choosing the model parameters such that the predicted distance-modulus curve closely follows its $\Lambda$CDM counterpart, we found that the corresponding Hubble-parameter curves still exhibit deviations from the $\Lambda$CDM prediction throughout the range $-0.1<\epsilon<+0.1$. Nevertheless, the corresponding diagrams show that the predicted curves pass through slightly more observational data points. For the same values of the model parameters, the latter theory predicts a larger age of the Universe than the $\Lambda$CDM model for both the quintaverse and the phantiverse scenarios. Furthermore, the $\gamma$-function remains positive throughout the cosmic evolution and increases from small values at high redshifts. In this theory, the effective {\sf DM} era is therefore associated with small values of the $\gamma$-function (see the panels of Fig.~\ref{fig9}).

\item
The kinematic diagnostics exhibit a similar overall pattern, with one important exception. In the cosmographic plane, all trajectories originate far from the $\Lambda$CDM matter fixed point (see the left panel of Fig.~\ref{fig10}). This behavior explains why the matter era in this theory is clearly distinguishable from its $\Lambda$CDM counterpart.

\end{itemize}

\appendix
\setcounter{equation}{0}
\renewcommand{\theequation}{A\arabic{equation}}

\section*{Appendix }\label{app}

This appendix collects several lengthy expressions that are referenced throughout the main text for completeness. For the coefficients $\mathfrak{T}_i$ appearing in eq.~(\ref{t2.31}) we obtain

\begin{align}\label{app1}
&\mathfrak{T}_0=-82944 \eta _2 \left(7575 \eta _1^4+26060 \eta _2 \eta _1^3-103006 \eta _2^2 \eta _1^2-172292 \eta _2^3 \eta _1+659455 \eta _2^4\right)\epsilon\\
&-62208 \eta _2 \left(-7575 \eta _1^4+53020 \eta _2 \eta _1^3+419814 \eta _2^2 \eta _1^2+536572 \eta _2^3 \eta _1+734105 \eta _2^4\right)\epsilon^2\nonumber\\
&+746496 \eta _2^2 \left(9885 \eta _1^3+14491 \eta _2 \eta _1^2-44493 \eta _2^2 \eta _1+38597 \eta _2^3\right)\epsilon^3\nonumber\\
&+279936 \eta _2^2 \left(-9885 \eta _1^3+111059 \eta _2 \eta _1^2+369953 \eta _2^2 \eta _1+359305 \eta _2^3\right)\epsilon^4\nonumber\\
&+839808 \eta _2^3 \left(-37665 \eta _1^2-10362 \eta _2 \eta _1+46331 \eta _2^2\right)\epsilon^5\nonumber\\
&-629856 \eta _2^3 \left(-12555 \eta _1^2+142006 \eta _2 \eta _1+200533 \eta _2^2\right)\epsilon^6\nonumber\\
&+3779136 \left(15585 \eta _1-2939 \eta _2\right) \eta _2^4\epsilon^7-708588 \left(15585 \eta _1-116789 \eta _2\right) \eta _2^4\epsilon^8\nonumber\\
&-40336371900 \eta _2^5\epsilon^9+6050455785 \eta _2^5\epsilon^{10},\nonumber\\
&\mathfrak{T}_1=-41472 \eta _2 \left(355 \eta _1^3+775 \eta _2 \eta _1^2+1353 \eta _2^2 \eta _1+7501 \eta _2^3\right)\epsilon\nonumber\\
&-31104 \eta _2 \left(-355 \eta _1^3+3033 \eta _2 \eta _1^2+11415 \eta _2^2 \eta _1+17811 \eta _2^3\right)\epsilon^2\nonumber\\
&+373248 \eta _2^2 \left(476 \eta _1^2+607 \eta _2 \eta _1+681 \eta _2^2\right)\epsilon^3+279936 \eta _2^2 \left(-238 \eta _1^2+2169 \eta _2 \eta _1+4449 \eta _2^2\right)\epsilon^4\nonumber\\
&-209952 \eta _2^3 \left(2967 \eta _1+1777 \eta _2\right)\epsilon^5+157464 \left(989 \eta _1-6025 \eta _2\right) \eta _2^3\epsilon^6+669851856 \eta _2^4\epsilon^7-125597223 \eta _2^4\epsilon^8,\nonumber\\
&\mathfrak{T}_2=5184 \eta _2 \left(81 \eta _1^2-2 \eta _2 \eta _1+305 \eta _2^2\right)\epsilon+3888 \eta _2 \left(-81 \eta _1^2+286 \eta _2 \eta _1+195 \eta _2^2\right)\epsilon^2\nonumber\\
&-23328 \eta _2^2 \left(71 \eta _1+95 \eta _2\right)\epsilon^3+8748 \left(71 \eta _1-55 \eta _2\right) \eta _2^2\epsilon^4+1180980 \eta _2^3\epsilon^5-295245 \eta _2^3\epsilon^6,\nonumber\\
&\mathfrak{T}_3=864 \eta _1 \eta _2-432 \eta _1^2+\Big(-432+3888 \epsilon ^2-5832 \epsilon ^3+2187 \epsilon ^4\Big)\eta _2^2.\nonumber
\end{align}

The following expressions give the explicit forms of the functions $f_i(X_3,X_5,X_6)$ and $g_i(X_3,X_5,X_6)$ appearing in eqs.~(\ref{t2.39.a}) and~(\ref{t3.25}). For brevity, we use the notation $X=X_3$, $Y=X_5$ and $Z=X_6$.

\begin{align}\label{app2}
&f_3(X,Y,Z)=\frac{1}{768 \eta _2 U_{\text{XYZ}}^4 Y}\Bigg\{32 \eta _2 U^3{}_{\text{XYZ}} X \bigg[12 Y Z^2 (3 U_{\text{XYZ}}+2 Y)-U_{XY}^2 \left(R (27 Z-4 X)+18 Y (8 U_{\text{XYZ}}-Z)+48 Y^2\right)\bigg]\nonumber\\
&-U_{\text{XYZ}}^2 \left(X^2-1\right)\big(\mathcal{H}_0+9\mathcal{H}_1Y+2\mathcal{H}_2Y^2+18\mathcal{H}_3Y^3+\mathcal{H}_4Y^4\big)\Bigg\},\\
&\mathcal{H}_0=-32 \eta _2 X (4 X-27 Z)U_{XZ}^2 ,\nonumber\\
&\mathcal{H}_1=8 \eta _1 Z \left(38 U_{\text{XYZ}} X+27 X^2-27\right)+64 \eta _2 U_{\text{X}}^2 \big(27 R \epsilon  (3 \epsilon -4)+8 X\big)\nonumber\\
&+6 \eta _2 Z \bigg[2 U_{\text{XYZ}} X \Big(19 \epsilon  (4-3 \epsilon )+60\Big)+9 X^2 \Big(13 \epsilon  (3 \epsilon -4)+12\Big)-9 \Big(13 \epsilon  (3 \epsilon -4)+12\Big)\bigg]\nonumber\\
&+Z^2 \bigg[27 U_{\text{XYZ}} \Big(\eta _2 \big(67 \epsilon  (4-3 \epsilon )+28\big)-12 \eta _1\Big)-512 \eta _2 X\bigg]+54 Z^3 \Big(4 \eta _1+\eta _2 \big(13 \epsilon  (3 \epsilon -4)+12\big)\Big),\nonumber\\
&\mathcal{H}_2=-64 \eta _2 \Big[6 U_{\text{XYZ}} X+X^2 \big(162 \epsilon  (3 \epsilon -4)-325\big)+162 \big(\epsilon  (4-3 \epsilon )+2\big)\Big]+81 Z^2 \Big[20 \eta _1+\eta _2 \big(119 \epsilon  (4-3 \epsilon )+220\big)\Big],\nonumber\\
&-54 Z \bigg[3 R \Big(36 \eta _1+\eta _2 \big(37 \epsilon  (3 \epsilon -4)-116\big)\Big)+8 \eta _2 X\bigg],\nonumber\\
&\mathcal{H}_3=96 \eta _2 U_{\text{XYZ}} \big(9 \epsilon  (4-3 \epsilon )+4\big)-256 \eta _2 X+27 Z \Big(4 \eta _1+\eta _2 \big(13 \epsilon  (3 \epsilon -4)+12\big)\Big),\nonumber\\
&\mathcal{H}_4=-20736 \eta _2 \big(\epsilon  (3 \epsilon -4)-2\big),\nonumber
\end{align}

\begin{align}\label{app3}
&f_5(X,Y,Z)=-\frac{1}{768 \eta _2 U_{\text{XYZ}}^3}\Bigg\{\mathcal{I}_0+\mathcal{I}_1X+12 \mathcal{I}_2X^2+\mathcal{I}_3X^3\Bigg\}\\
&\mathcal{I}_0=15552 \eta _2 X^5 Y \epsilon  (3 \epsilon -4)+36 X^4 Y \bigg[64 \eta _2 Y-76 \eta _1 Z+\eta _2 Z \Big(57 \epsilon  (3 \epsilon -4)-196\Big)\bigg]\nonumber\\
&-96 \eta _2 U_{YZ}\Big[6 Y^3 \left(8 U_{\text{XYZ}}-Z\right)+Y^2 \left(9 Z U_{\text{XYZ}}+8 Z^2-16\right)+6 Y \left(2 Z^2 U_{\text{XYZ}}-8 U_{\text{XYZ}}+Z\right)-9 Z U_{\text{XYZ}}+16 Y^4\Big]\nonumber\\
&\mathcal{I}_1=1728 \eta _2 Y^5 \big[9 \epsilon  (3 \epsilon -4)-4\big]-128 \eta _2 U_{\text{XYZ}} U_Z\nonumber\\
&+4 Y^4 \bigg[81 Z \Big(36 \eta _1+\eta _2 \big(37 \epsilon  (3 \epsilon -4)-116\big)\Big)-32 \eta _2 \big(162 \epsilon  (3 \epsilon -4)-323\big) U_{\text{XYZ}}\bigg]\nonumber\\
&+2 Y^2 \bigg[Z^2 U_{\text{XYZ}} \Big(1620 \eta _1+\eta _2 \big(17756-9639 \epsilon  (3 \epsilon -4)\big)\Big)+128 \eta _2 \big(81 \epsilon  (3 \epsilon -4)-161\big) U_{\text{XYZ}}\nonumber\\
&-162ZU^2_Z\Big(36 \eta _1+\eta _2 \big(37 \epsilon  (3 \epsilon -4)-116\big)\Big)\bigg]\nonumber\\
&-243 Y U_Z \bigg[2 Z U_{\text{XYZ}} \Big(4 \eta _1+\eta _2 \big(13 \epsilon  (3 \epsilon -4)+12\big)\Big)+Z^2 \Big(12 \eta _1+\eta _2 \big(67 \epsilon  (3 \epsilon -4)-28\big)\Big)+64 \eta _2 \epsilon  (4-3 \epsilon )\bigg]\nonumber\\
&+27 Y^3\bigg[18 Z U_{\text{XYZ}} \Big(4 \eta _1+\eta _2 \big(13 \epsilon  (3 \epsilon -4)+12\big)\Big)+28 \eta _2 \big(9 \epsilon  (4-3 \epsilon )+2\big) \nonumber\\
&+Z^2 \left(108 \eta _1+\eta _2 (1179 \epsilon  (3 \epsilon -4)-508)\right)\bigg],\nonumber\\
&\mathcal{I}_2=4 \eta _2\Big[Y \big(U_Y^2 (159 Z-80 Y)+Z^2 (64 Y-147 Z)\big)+6 U_{\text{XYZ}} \big(16 Y^3+3 Y^2 Z+4 Y (Z^2-4)-3 Z\big)\bigg]\nonumber\\
&+57 Y Z U_{\text{YZ}} \big(4 \eta _1+3 \eta _2 \epsilon  (4-3 \epsilon )\big),\nonumber\\
&\mathcal{I}_3=972  Y Z \big(2 U_{\text{XYZ}}+3 (4 Y+Z)\big)\eta _1\nonumber\\
&+\bigg[U_{\text{XYZ}} \Big(128 Y^2 \big(162 \epsilon  (4-3 \epsilon )+323\big)+486 Y Z \big(13 \epsilon  (3 \epsilon -4)+12\big)+128\Big)\nonumber\\
&+27 Y \Big(128 Y^2 \big(9 \epsilon  (3 \epsilon -4)-2\big)+12 Y Z \big(37 \epsilon  (3 \epsilon -4)-116\big)+9 Z^2 \big(131 \epsilon  (3 \epsilon -4)-28\big)+1152 \epsilon  (4-3 \epsilon )\Big)\bigg]\eta_2,\nonumber
\end{align}

\begin{align}\label{app4}
&f_6(X,Y,Z)=-\frac{Z}{768 \eta _2 U_{\text{XYZ}}^3}\Bigg\{\mathcal{J}_0+\mathcal{J}_1X+12\mathcal{J}_2X^2+\mathcal{J}_3X^3+34\mathcal{J}_4X^4\Bigg\},\\
&\mathcal{J}_0=15552 \eta _2 X^5 \epsilon  (3 \epsilon -4)-96 \eta _2 U_{\text{YZ}}^2 \big(6 Y^2 (8 U_{\text{XYZ}}-Z)+Y (9 Z U_{\text{XYZ}}-8 U^2_{\text{Z}} )-12 U^2_{\text{Z}} U_{\text{XYZ}}+16 Y^3\big),\nonumber\\
&\mathcal{J}_1=1728 \eta _2 Y^4 \big(9 \epsilon  (3 \epsilon -4)-4\big)+4Y^3\bigg[81 Z \Big(36 \eta _1+\eta _2 \big(37 \epsilon  (3 \epsilon -4)-116\big)\Big)-32 \eta _2 \big(162 \epsilon  (3 \epsilon -4)-323\big) U_{\text{XYZ}}\bigg]\nonumber\\
&+2 Y \bigg[Z^2 U_{\text{XYZ}} \Big(1620 \eta _1+\eta _2 \big(17756-9639 \epsilon  (3 \epsilon -4)\big)\Big)+64 \eta _2 \big(162 \epsilon  (3 \epsilon -4)-323\big) U_{\text{XYZ}}\nonumber\\
&-162 Z U_Z^2 \Big(36 \eta _1+\eta _2 \big(37 \epsilon  (3 \epsilon -4)-116\big)\Big)\bigg]\nonumber\\
&-243 U_Z^2 \bigg[2 Z U_{\text{XYZ}} \Big(4 \eta _1+\eta _2 \big(13 \epsilon  (3 \epsilon -4)+12\big)\Big)+Z^2 \Big(12 \eta _1+\eta _2 \big(67 \epsilon  (3 \epsilon -4)-28\big)\Big)+64 \eta _2 \epsilon  (4-3 \epsilon )\bigg]\nonumber\\
&+27 Y^2 \bigg[\eta _2 \Big(18 Z \big(13 \epsilon  (3 \epsilon -4)+12\big) U_{\text{XYZ}}+Z^2 \big(1179 \epsilon  (3 \epsilon -4)-508\big)+128 \big(9 \epsilon  (4-3 \epsilon )+2\big)\Big)\nonumber\\
&+36 \eta _1 Z \left(2 U_{\text{XYZ}}+3 Z\right),\nonumber\\
&\mathcal{J}_2=3 U^2_Z \Big(-32 \eta _2 U_{\text{XYZ}}+76 \eta _1 Z+\eta _2 Z \big(57 \epsilon  (4-3 \epsilon )+196\big)\Big)+8 \eta _2 Y (9 Z U_{\text{XYZ}}+32 Z^2-32)+320 \eta _2 Y^3\nonumber\\
&+3 Y^2 \Big(128 \eta _2 U_{\text{XYZ}}-76 \eta _1 Z+\eta _2 Z \big(57 \epsilon  (3 \epsilon -4)-212\big)\Big),\nonumber\\
&\mathcal{J}_3=4 Y \bigg[81 Z \Big(36 \eta _1+\eta _2 \big(37 \epsilon  (3 \epsilon -4)-116\big)\Big)-32 \eta _2 \big(162 \epsilon  (3 \epsilon -4)-323\big) U_{\text{XYZ}}\bigg]\nonumber\\
&+3456 \eta _2 Y^2 \big(9 \epsilon  (3 \epsilon -4)-2\big)+243\bigg[2 Z U_{\text{XYZ}} \Big(4 \eta _1+\eta _2 \big(13 \epsilon  (3 \epsilon -4)+12\big)\Big)+128 \eta _2 \epsilon  (4-3 \epsilon )\nonumber\\
&+Z^2 \Big(12 \eta _1+\eta _2 \big(131 \epsilon  (3 \epsilon -4)-28\big)\Big)\bigg],\nonumber\\
&\mathcal{J}_4=64 \eta _2 Y-76 \eta _1 Z+\eta _2 Z \big(57 \epsilon  (3 \epsilon -4)-196\big).
\end{align}

\begin{align}\label{app5}
&g_3(X,Y,Z)=\frac{1}{768 \eta_2 Y U_{XYZ}^2}\left[\eta _1 T_{31}+\eta _2 T_{32}+9 \epsilon  (4-3 \epsilon) \eta _2T_{33}\right],\\
&T_{31}=12 Y Z U_X^2 \left[U_{\text{XYZ}} (4 X+180 Y-81 Z)+162 U_{\text{XY}}^2-54 Z (Y+3 Z)\right],\nonumber\\
&T_{32}=4\Bigg\{8 X U_{XYZ} \bigg[U_{XY}^2 \Big((18 Y+1) U_{XYZ}+4 X+9 (2 Y-3) Z-48 Y (Y+3)\Big)+12 Y Z^2 \left(-U_{XYZ}+2 Y+3\right)\bigg]\nonumber\\
&-1152 X Y U_X^4+U_X^2\bigg[8 X U_{XZ}^2 (4 X-27 Z)-3 Y U_{XYZ} \left(-64 X Y-92 X Z+576 Y^2+3204 Y Z+243 Z^2\right)\nonumber\\
&+2 Y \Big(81 U_Y^2 (64 Y+9 Z)+9 Z^2 (64 X-531 Y)+27 X Z (4 Y-27 X)+16 X Y (36 Y-325 X)-729 Z^3\Big)\bigg]\Bigg\},\nonumber\\
&T_{33}= -Y U_X^2 \bigg[18 \Big(U_Y^2 (39 Z-128 Y)+X^2 (128 Y-39 Z)+3 Z^2 (41 Y-13 Z)\big)\nonumber\\
&+U_{\text{XYZ}} \Big(-1728 U_X^2-4 X Z+1728 Y^2+1356 Y Z+1701 Z^2\Big)\bigg],
\end{align}

\begin{align}\label{app6}
&g_5(X,Y,Z)=\frac{1}{768 \eta_2 U_{XYZ}^3}\left[\eta _1 T_{31}+\eta _2 T_{32}+9 \epsilon  (4-3 \epsilon) \eta _2T_{33}\right],\\
&T_{51}=-12 X Y Z \bigg[U_{XY}^2 \big(81 Z-4 (X+45 Y)\big)+Z \big(81 U_Z^2+4 Z (X+45 Y)-81\big)\nonumber\\
&+54 U_{XYZ} \big(-3 U_{XY}^2-3 U_Z^2+Y Z+3\big)\bigg),\nonumber\\
&T_{52}=-4\Bigg\{3Y\bigg[4 X U_Y^2 \big(X (Z-80 Y)+9 Y (16 Y+89 Z)\big)-16 U_{YZ}^2 \big(-8 Y U_Y^2+Y (-3 Y Z+4 Z^2+8)+3 Z\big)\nonumber\\
&+X \Big(243 Z^2 U_Z^2+4 X^3 (48 Y+11 Z)-9 X^2 (64 Y^2+356 Y Z+27 Z^2)\nonumber\\
&+4 X Z^2 (64 Y+11 Z)-9 Y Z^2 (91 Y+356 Z)\Big)\bigg]+2 U_{XYZ} \bigg[36 U_{XYZ}^2 U_Y^2 (16 Y+3 Z)+144 Y U_{YZ}^2 (U_Z^2-1)\nonumber\\
&X \Big(-U_Z^2 (144 X Y+729 Y Z+16)-729 Y Z (U_{XY}^2-1)\nonumber\\
&+16 \big((323 X^2-322) Y^2+X^2+9 X Y+323 Y^4\big)+4763 Y^2 Z^2\Big)\bigg]\Bigg\},\nonumber\\
&T_{53}=XY\bigg[1728 U_{\text{XY}}^4+Z^2 (-4 X Z+1356 Y Z+3429)-1728\nonumber\\
&+U_{XY}^2 \big(18 U_{XYZ} (128 Y-39 Z)+Z (4 X-1356 Y-3429 Z)\big)\nonumber\\
&-54 Z U_{XYZ} (13 U_Z^2+41 Y Z-13)+27 (63 U_Z^2+1)U_Z^2\bigg],\nonumber\\
\end{align}

\begin{align}\label{app7}
&g_6(X,Y,Z)=\frac{1}{768 \eta_2 U_{XYZ}^3}\left[\eta _1 T_{61}+\eta _2 T_{62}+9 \epsilon  (4-3 \epsilon) \eta _2T_{63}\right],\\
&T_{61}=T_{51},\nonumber\\
&T_{62}=T_{52}-1536 Y^2 U_{\text{YZ}}^2,\nonumber\\
&T_{63}=X Z \Big[4 X Z (U_{XY}^2+339 X Y-1)-3429 Z^2 (U_{XY}^2-1)+1356 Y Z^3\nonumber\\
&+U_Z^2 (1701 U_Z^2+4 X Z+27)-18 U_{XYZ} \big(U_{XY}^2 (39 Z-128 Y)+3 Z (13 U_Z^2+41 Y Z-13)\big)\nonumber\\
&+1728 (U_{XY}^4-1)-1356 Y Z U_Y^2.
\end{align}

Where, for the sake of brieffness we have defined 

\begin{align}\label{App5}
&U_{XYZ}=\sqrt{1-X^2-Y^2-Z^2},\\
&U_{XY}=\sqrt{1-X^2-Z^2},\\
&U_{X}=\sqrt{1-X^2},
\end{align}

and similar relations.



\end{document}